\documentclass[aps,prx,longbibliography,showpacs,twocolumn,superscriptaddress,amsmath,amssymb,verbatim]{revtex4-2}
\usepackage{amsmath}
\usepackage{epsfig}
\usepackage{graphicx}
\usepackage[outdir=./PDFGraph/]{epstopdf}
\usepackage{natbib}
\usepackage{array}
\usepackage{setspace}
\usepackage{mathrsfs}
\usepackage{amssymb}
\usepackage{soul}
\usepackage[usenames, dvipsnames]{color}
\usepackage[breaklinks=true,colorlinks,linkcolor=blue]{hyperref}
\usepackage{float}
\usepackage[normalem]{ulem}

\usepackage[yyyymmdd]{datetime}

\newcommand{\SPhide}[1]{{}}

\begin{document}

\title{
Unusual spin dynamics in the low-temperature magnetically ordered state of 
Ag$_{3}$LiIr$_{2}$O$_{6}$}

\author{Atasi Chakraborty} 
\thanks{Equal contribution authors} 
\affiliation{School of Physical Sciences, Indian Association for the Cultivation of Science, Jadavpur, Kolkata 700032, India}

\author{Vinod Kumar} 
\thanks{Equal contribution authors} 
\affiliation{Department of Physics, Indian Institute of Technology Bombay, Powai, Mumbai 400076, India}

\author{Sanjay Bachhar} 
\affiliation{Department of Physics, Indian Institute of Technology Bombay, Powai, Mumbai 400076, India}

\author{N. B\"{u}ttgen}
\affiliation{Experimentalphysik V, Elektronische Korrelationen und Magnetismus, Institut f\"{u}r Physik, Universit\"{a}t Augsburg, 86135 Augsburg, Germany}

\author{K. Yokoyama}
\affiliation{ISIS Pulsed Neutron and Muon Source, STFC Rutherford Appleton Laboratory,	Harwell Campus, Didcot, Oxfordshire OX110QX, UK}

\author{P. K.  Biswas}
\affiliation{ISIS Pulsed Neutron and Muon Source, STFC Rutherford Appleton Laboratory,	Harwell Campus, Didcot, Oxfordshire OX110QX, UK}

\author{V. Siruguri}
\affiliation{UGC-DAE-Consortium for Scientific Research Mumbai Centre, Bhabha Atomic Research Centre, Mumbai 400085, India}

\author{Sumiran Pujari}
\affiliation{Department of Physics, Indian Institute of Technology Bombay, Powai, Mumbai 400076, India}

\author{I. Dasgupta}
\affiliation{School of Physical Sciences, Indian Association for the Cultivation of Science, Jadavpur, Kolkata 700032, India}

\author{A.V. Mahajan}
\thanks{corresponding author}
\affiliation{Department of Physics, Indian Institute of Technology Bombay, Powai, Mumbai 400076, India}

\date{\today}

\begin{abstract}
Recently, there have been contrary claims of Kitaev spin-liquid behaviour and ordered 
behavior in the honeycomb compound Ag$_3$LiIr$_2$O$_6$ based on various 
experimental signatures. Our investigations on this system reveal a 
low-temperature ordered state with persistent dynamics down to the lowest temperatures. 
Magnetic order is confirmed by clear oscillations in the muon spin 
relaxation ($\mu$SR) time spectrum  below 9 K till 52 mK.  
Coincidentally in $^7$Li nuclear magnetic resonance, %(NMR), 
a wipe-out of the %NMR 
signal is observed below $\sim$ 10 K which again strongly indicates 
magnetic order in the low temperature regime.  
This is supported by our density functional theory 
calculations which show an appreciable 
Heisenberg exchange term in the spin Hamiltonian 
that favors magnetic ordering.
The $^7$Li shift and spin-lattice relaxation rate also show anomalies at $\sim$ 50 K.
They are likely related to the onset of dynamic magnetic correlations,
but their origin is not completely clear.
Detailed analysis of our $\mu$SR data is consistent with a co-existence of 
incommensurate N\'eel and striped environments. 
A significant and undiminished dynamical relaxation rate ($\sim 5$ MHz) as seen in $\mu$SR 
deep into the ordered phase indicates enhanced quantum fluctuations in the ordered state. 
\end{abstract}
\maketitle

\section{Introduction}
\label{sec:intro}

Kitaev's seminal proposal of bond-dependent magnetic interactions 
stabilizing a novel $Z_2$ spin-liquid ground state with Majorana 
excitations, followed by the important material-specific
advance of Jackeli and Khaliullin~\cite{Jackeli_2009}
has triggered significant experimental effort to 
synthesize such materials. They advocated
honeycomb lattice structures of $4d$/$5d$ element based oxides 
with edge-sharing oxygen octahedra and strong spin-orbit 
coupling as having the necessary ingredients to host 
the Kitaev model~\cite{Chaloupka_K_2010}.
Several promising candidates with such layered 
honeycomb structure have since been investigated: 
Na$_{2}$IrO$_{3}$~\cite{Chaloupka_Z_2013, Hou_F_2018, Katukuri_K_2014}, $\alpha$-Li$_{2}$IrO$_{3}$ 
(as also its three dimensional 
polymorphs)~\cite{Katukuri_T_2016, Williams_I_2016}, 
and $\alpha$-RuCl$_{3}$~\cite{Do_M_2017,Hou_U_2017}.
However, 
%experiments and theoretical calculations 
it has been revealed 
that these materials order 
magnetically~\cite{Gretarsson_C_2013,Manni_E_2014,Hermann_P_2019,Mazin_O_2013,Choi_S_2012,Winter_C_2016,Hou_U_2017} 
due to the presence of Heisenberg and other non-Kitaev terms, 
and the fingerprint of the Kitaev interactions may only be 
realised  either at higher temperatures or under 
application of a magnetic field. 

In this family of materials a new addition has been 
H$_{3}$LiIr$_{2}$O$_{6}$~\cite{Bette_S_2017}, where all 
of the interlayer Li$^{+}$ ions of $\alpha$-Li$_{2}$IrO$_{3}$ (LIO) 
are replaced by  H$^{+}$ ions, retaining the LiIr$_{2}$O$_{6}$ planes. 
Various measurements have confirmed the absence of magnetic ordering 
down to 0.05~K in H$_{3}$LiIr$_{2}$O$_{6}$~\cite{Kitagawa_A_2018}
which has been argued to be a spin-orbit entangled quantum 
spin-liquid~\cite{Kitagawa_A_2018}. To complicate matters further, 
x-ray diffraction  (XRD) has revealed   
the presence of stacking faults between the honeycomb planes~\cite{Bette_S_2017},
and the low temperature behavior was attributed to local 
moments induced by these defects. Theoretical \textit{ab-initio}
calculations have also shown for these systems 
that although the bond-dependent Kitaev interactions are significant, 
the Heisenberg and other non-Kitaev terms are not negligible. 
It has been suggested that these systems lie close to 
the tricritical point between ferromagnetic, zigzag and incommensurate 
spiral order resulting in the absence of magnetic order~\cite{Li_R_2018}. 
Calculations further reveal that the interlayer O-H-O geometry as well 
as lack of hydrogen order also strongly influence the Kitaev and other exchange 
interactions having strong impact on its magnetic properties.\cite{Li_R_2018, Yadav_S_2018}

Very recently, the compound Ag$_{3}$LiIr$_{2}$O$_{6}$ (ALIO) has
been synthesized~\cite{Bette_C_2019}, where 
proximate Kitaev spin-liquid physics has been claimed based on 
scaling behaviour of various thermodynamic quantities in the presence 
of quenched disorder and a two step release of magnetic entropy.
Replacement of the lighter H$^{+}$ ions 
by the heavier Ag$^{+}$ ions leads to an 
increase in the inter-layer separation 
which can significantly influence the various magnetic exchange interactions. 
Estimates of the magnetic interactions using 
{\em ab-initio} study are currently lacking for this system.

We have been working on the hexagonal Ag$_{3}$LiM$ _{2} $O$ _{6} $ (M = Mn, Ru, Ir) 
system with the intention of developing a comprehensive understanding of this
honeycomb system.
The Mn-based ($3d^{3} $ or $S = 3/2$) system exhibits long-range order below about 50 K. 
There is also evidence of Berezinskii-Kosterlitz-Thouless behavior from an 
analysis of the electron spin resonance line broadening data near the 
transition temperature~\cite{Kumar_S_2019}.
The Ru-based system ($4d^{4} $ which might be $S = 1$ or $J = 0$) was expected to be 
a possible candidate for excitonic magnetism~\cite{Khaliullin_E_2013}.
Our investigations however revealed that spin-orbit coupling might 
not be significant in this case.  
But the expected long-range ordered state was not seen 
in spite of the apparently unfrustrated geometry of the honeycomb lattice. 
Rather, the local moments were found to be on the borderline of being dynamic and 
static at low temperatures, based on 
muon spin relaxation ($\mu$SR) data~\cite{Kumar_U_2019}.
Such behavior is likely driven by higher order bi-quadratic or 
ring exchange terms in the Hamiltonian over and above the usual Heisenberg couplings. 

Continuing our investigations in this series of systems, we focus in this paper 
on the Ir analog containing Ir$^{4+}$ ions ($J_{\text{eff}} = 1/2$) with the 
purpose of looking for possible Kitaev spin-liquid physics due to
the enhanced spin-orbit coupling of the Ir moments. 
Our experimental %(NMR and $\mu$SR) 
and theoretical results as itemized below 
are however quite far from such expectations: 
\begin{itemize}
\item We observe clear oscillations in $\mu$SR relaxation data below $\sim 9$K
providing strong evidence for magnetic order below this temperature.
Analysis of the muon data in the ordered state, 
complemented with density function theory (DFT) simulations of the muon stopping site 
points towards the co-existence of 
incommensurate N\'eel and stripe ordered magnetic domains. 

\item We also provide theoretical estimates for the various magnetic 
interactions in ALIO via DFT-based computations. We find 
the ratio of the nearest neighbor Kitaev exchange ($K_{1}$) to the 
Heisenberg exchange ($J_{1}$) to be in the range 
$\vert \frac{K_{1}}{J_{1}} \vert$ $\sim$ 2.0-3.5, 
placing the ALIO system far from the pure Kitaev limit,
and closer to the phase boundaries
between stripe, N\'eel and 120$^{\circ}$  
order of the phase diagram~\cite{Rau_G_2014}.

\item We observe two anomalies in the $^7$Li 
nuclear magnetic resonance (NMR) shift variation with temperature.  
The first one that is present 
at $T \sim$ 10 K clearly signifies the onset of magnetic long-range order 
as seen from the so-called wipe-out of the NMR signal, which agrees with our conclusions
from muon data.

\item 
The second anomaly is a broad maximum at about 50 K in the temperature
variation of the $^7$Li NMR line shift (which tracks the intrinsic spin susceptibility) 
and is reminiscent of such features in quasi-one and quasi-two dimensional 
Heisenberg antiferromagnetic systems due to short-range magnetic correlations.  
The $^7$Li NMR 1/$T_1$ also has a maximum at $T \sim$ 50 K. 
Together with the observed progressive loss of the NMR intensity
at this temperature, this could be
signifying dynamic short-range magnetic correlations similar to those 
seen in CeCu$_2$Si$_2$~\cite{Nakamura_1992,Kitaoka_1991}.

\item
Finally, we also find 
a large value of the muon relaxation rate ($\sim 5$ MHz)
that remains essentially flat and undiminished
deep into the ordered phase, i.e. down to 52 mK which is about
$1/200^{\text{th}}$ of the ordering temperature of $\sim 10$ K.
This is quite striking and noteworthy, 
rather reminiscent of spin-liquid behavior~\cite{ZhuZiHao2020} in spite of 
the unambiguous evidence for
magnetic order mentioned above (i.e., clear oscillations in $\mu$SR relaxation data.)
We interpret this as a signature of persistent spatio-temporal
fluctuations of the  N\'eel and stripe ordered domains,
possibly driven by quantum effects given our theoretical estimates of
the various magnetic exchange energy scales in ALIO.

\end{itemize}

Our results are in sharp contrast with a recent report 
on the same compound~\cite{Bahrami_T_2019}.
This report was based on the low-temperature scaling behaviors as stated before, and
in particular, on an apparent two step release of magnetic entropy 
suggesting ALIO may be a proximate Kitaev spin-liquid. 
We argue that the (extracted) magnetic specific heat in the high-temperature regime
is very uncertain.  
This is simply due to the overwhelming contribution of the lattice
to the total specific heat, 
especially in the high-temperature region ($T \gtrsim$ 30 K).  
Consequently, the inference of the high-temperature peak 
(position and magnitude) and that of a two-step entropy release 
is rather insecure, let alone ascribing it to Majorana excitations. 
We note here that an even more recent report~\cite{Bahrami2020} has found
evidence for magnetic ordering in a cleaner batch of samples in line with our observations.

The remainder of the paper is organised as follows: 
we start by giving the details on the structure of ALIO and relevant technical details
on measurements and theoretical methods in Sec.~\ref{sec:technical}. 
We next present our main pieces of experimental evidence that establish
a low-temperature magnetic ordered state coming from
$\mu$SR and NMR data in Sec.~\ref{sec:NMR_muSR}.
This is followed by a presentation of our theoretical estimates for the
various magnetic exchange couplings based on
DFT in Sec.~\ref{sec:DFT}.
Sec.~\ref{sec:bulk_probes} is devoted to a detailed discussion of our observations using 
bulk probes (heat capacity and susceptibility) and the high temperature 
anomaly in NMR vis-a-vis long-range magnetic 
order below 10 K as inferred from our observations and computation, 
versus Kitaev spin-liquid scenario as in Ref.~\cite{Bahrami_T_2019}. 
Concluding remarks are given in Sec.~\ref{sec:conclu}.

\section{Structure and technical details}
\label{sec:technical}
\begin{figure*}
	\includegraphics[width=0.8\linewidth]{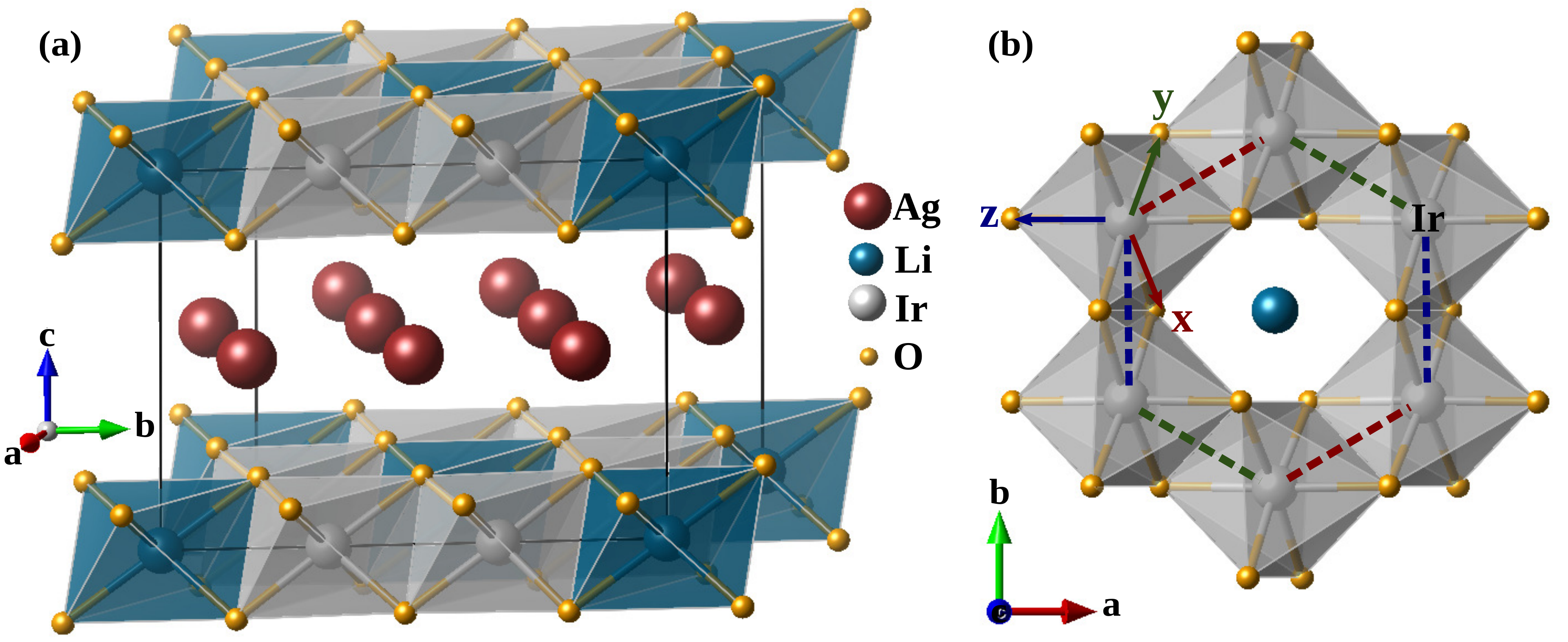}
	\caption{(a) shows the unit cell of Ag$_3$LiIr$_2$O$6$. The edge shared
		IrO$_6$ honeycomb network is shown (b). 
The $x, y, z$ local axes point toward the transition metal (Ir) 
to ligand (O) direction. X (red dotted line), Y (green dotted line) and 
Z (blue dotted line) type Ir-Ir nearest-neighbor bonds are perpendicular to 
the chosen $x, y, z$ local axes respectively.
}
	\label{strucFig}
\end{figure*}

ALIO crystallizes in base centered monoclinic symmetry having space
group C2/m. The crystal structure is shown in Fig.~\ref{strucFig}(a).
The magnetic building block consists of Ir  in an octahedral 
environment with nearest neighbor ($nn$) oxygen ligands.
The IrO$_6$ octahedra form an edge shared honeycomb geometry 
in the a-b plane containing Li ions at the center (see Fig.~\ref{strucFig}(b)).
The honeycomb layers in ALIO are identical to its parent 
compound, $\alpha-$LIO but the chemical bonds
between the layers are modified. The interlayer Li atoms 
in $\alpha-$LIO are octahedrally coordinated with
six oxygens in the two adjacent O$_6$ honeycomb layers, 
whereas the Ag atoms in ALIO are linearly connected to two 
oxygens in neighboring layers making a 180$^0$ O-Ag-O bond-angle.

Polycrystalline samples of ALIO were prepared by a two-step process as described in 
the supplementary information (SM)~\cite{SI}.
The resulting product was  Ag$_3$LiIr$_2$O$_6$ as verified by lab x-ray 
diffraction measurements using a PANalytical X'Pert PRO diffractometer 
using Cu-K${\alpha}$ radiation ($\lambda=1.54182$\AA). Small amounts of 
residual Ag and $\alpha$-Li$_2$IrO$_3$ were detected in the x-ray diffraction pattern.

The magnetization measurements have been performed in a Quantum 
Design SQUID Vibrating Sample Magnetometer in the temperature 
range 2-400\,K and in applied fields ranging from 0\,Oe to 70\,kOe.  
The heat capacity measurements have been done in a Quantum 
Design PPMS in the temperature range 2-250\,K, in various field 
values in the range 0-90\,kOe.  
%Muon spin relaxation 
$\mu$SR measurements were carried out 
using the MUSR spectrometer at the  ISIS Neutron and Muon Source 
at the STFC Rutherford Appleton Laboratory in the UK. The powder 
sample was loaded on a silver sample holder to minimize the 
background signal. The holder was then mounted on a dilution 
refrigerator insert and a standard cryostat stick for measuring 
temperatures ranging from 50 mK up to 150 K.
$^7$Li 
%nuclear magnetic resonance 
NMR measurements  have been 
performed in a fixed field of 93.954\,kOe, using a Tecmag 
spectrometer in a continuous flow cryostat in the temperature 
range of 4-300 K.  Measurements have also been performed in a 
swept field magnet down to 1.5 K at various frequencies (and 
therefore, fields). From our measurements, we obtained 
$^7$Li NMR spectra, spin-lattice (1/$T_1$) and 
spin-spin (1/$T_2$) relaxation  rates as a function of 
temperature in various fields. 

The first-principles electronic structure calculations in the framework of 
density functional theory (DFT) are carried out  within the generalized 
gradient approximation (GGA) for the exchange-correlation functional following
the Perdew-Burke-Ernzerhof prescription. We have employed the plane-wave 
basis as implemented within the Vienna Ab initio Simulation Package 
(VASP)~\cite{Kresse_A_1993,Kresse_E_1996} with projector augmented
wave potentials~\cite{Blochl_P_1994,Kresse_F_1999} as well as in the 
$N^{\text{th}}$-order muffin-tin orbital (NMTO) and
linear muffin-tin orbital (LMTO) basis sets as implemented 
in the STUTTGART code~\cite{Andersen_M_2000}.
The consistency between the two sets of calculations in two choices 
of basis sets is cross-checked in terms of band
structure, density of states etc. The VASP calculations are done with 
usual values of Coulomb correlation $U$~\cite{Anisimov_B_1991} and 
Hund’s coupling ($J_{H}$) chosen for Ir with $U_{eff}(\equiv U-J_{H})=1.5$ eV 
in Dudarev scheme~\cite{Dudarev_E_1998}. The details of the VASP, LMTO 
and NMTO calculations are described in the SM~\cite{SI}.

\section{The low temperature ordered state of ALIO} %probed by $\mu$SR and NMR}
\label{sec:NMR_muSR}

\subsection{$\mu$SR}
\label{subsec:muSR}

\begin{figure}[t]
	\includegraphics[width=1.0\columnwidth,clip=true,trim=20mm 25mm 10mm 20mm]{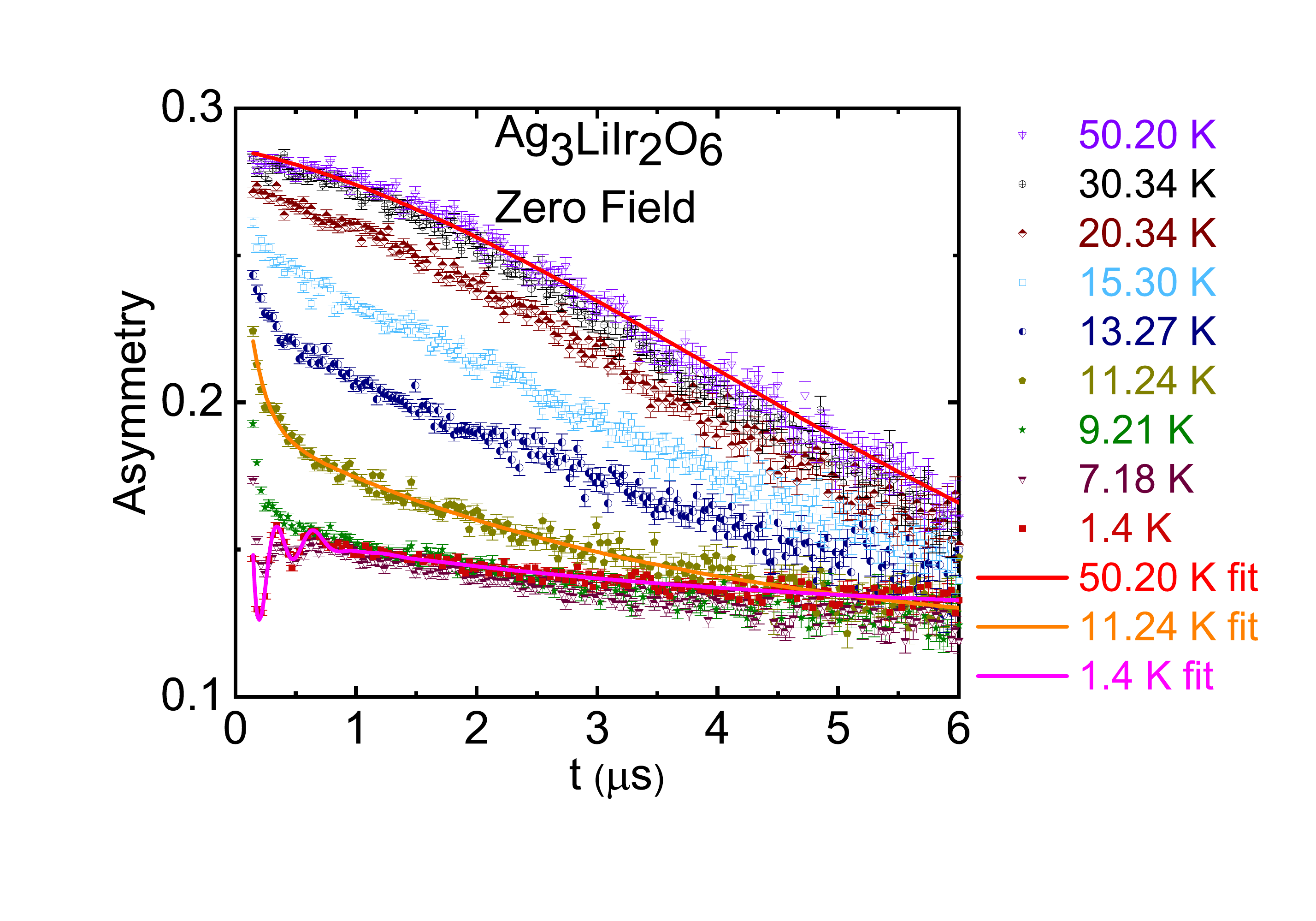}
	\caption{Variation of the muon asymmetry with time is shown at selected temperatures. Clear oscillations are seen below about 9 K indicative of long-range magnetic order. Fits at some representative temperatures are shown as explained in the text.
%\SP{too much black due to filled markers perhaps. May be hollow markers might be better?}
} 
	\label{musr-asymmetry}
\end{figure}

The depolarisation of the muons as a function of temperature is 
shown in Fig.~\ref{musr-asymmetry}. At high temperatures, a 
slow (Gaussian-like) decay of the muon polarisation is seen whereas 
below about 20 K, a faster (exponential-like) decay is discernible 
which gets even faster with a decrease in temperature.  Finally, below 
about 9 K, clear oscillations in the muon asymmetry as a function 
of time are seen.  Fits of the time decay of the muon asymmetry at 
various temperatures have then been carried out to obtain the 
variation of the local moment dynamics with temperature.

%%%%%%%%%%%%%%%%%%%%%%%%%%%%%%%%%%%%%%%%%%%%%%%%%%%%%%%%%%%%%%%%%%%%%%
%%%%%%%%%%%%%%%%%%%%%%%%%%%%%%%%%%%%%%%%%%%%%%%%%%%%%%%%%%%%%%%%%%%%%%%%%%%%%%%%%%%%%%%%%%%%
We find that at temperatures above 15 K, the data are well fit 
to a product of a static Kubo-Toyabe function with an 
exponential in addition to a constant background $A_0$ \textit{i.e.}, 
$\mathit{A\mathrm{(}t\mathrm{)}}=A_{2}G_{KT}(\Delta,t)e^{-\lambda _2(T)t}+A_{0}$.  
Here, $G_{KT}(\Delta,t)$ is the Kubo-Toyabe function which models the 
relaxation of muons in a Gaussian distribution of magnetic fields 
from nuclear moments.  From these fits we obtain the field 
distribution $\Delta$ to be about 1.6 Oe. This value is typical 
of nuclear dipolar fields at the muon site, in the present case 
arising from $^{107, 109}$Ag, $^{6, 7}$Li, and $^{191, 193}$Ir 
nuclei.  The exponential term $e^{-\lambda _2(T)t}$ arises from 
the relaxation due to fluctuations of the electronic local 
moments.  This relaxation rate $\lambda _2(T)$ is small at high 
temperatures and gradually increases as the local moment 
fluctuation rate gets smaller (see Fig.~\ref{musr-lambda}).  
We notice a sharper increase of $\lambda _2(T)$ below about 20 K.  

\begin{figure}[t]
	\includegraphics[width=1.0\columnwidth,clip=true,trim=0mm 0mm 20mm 15mm]{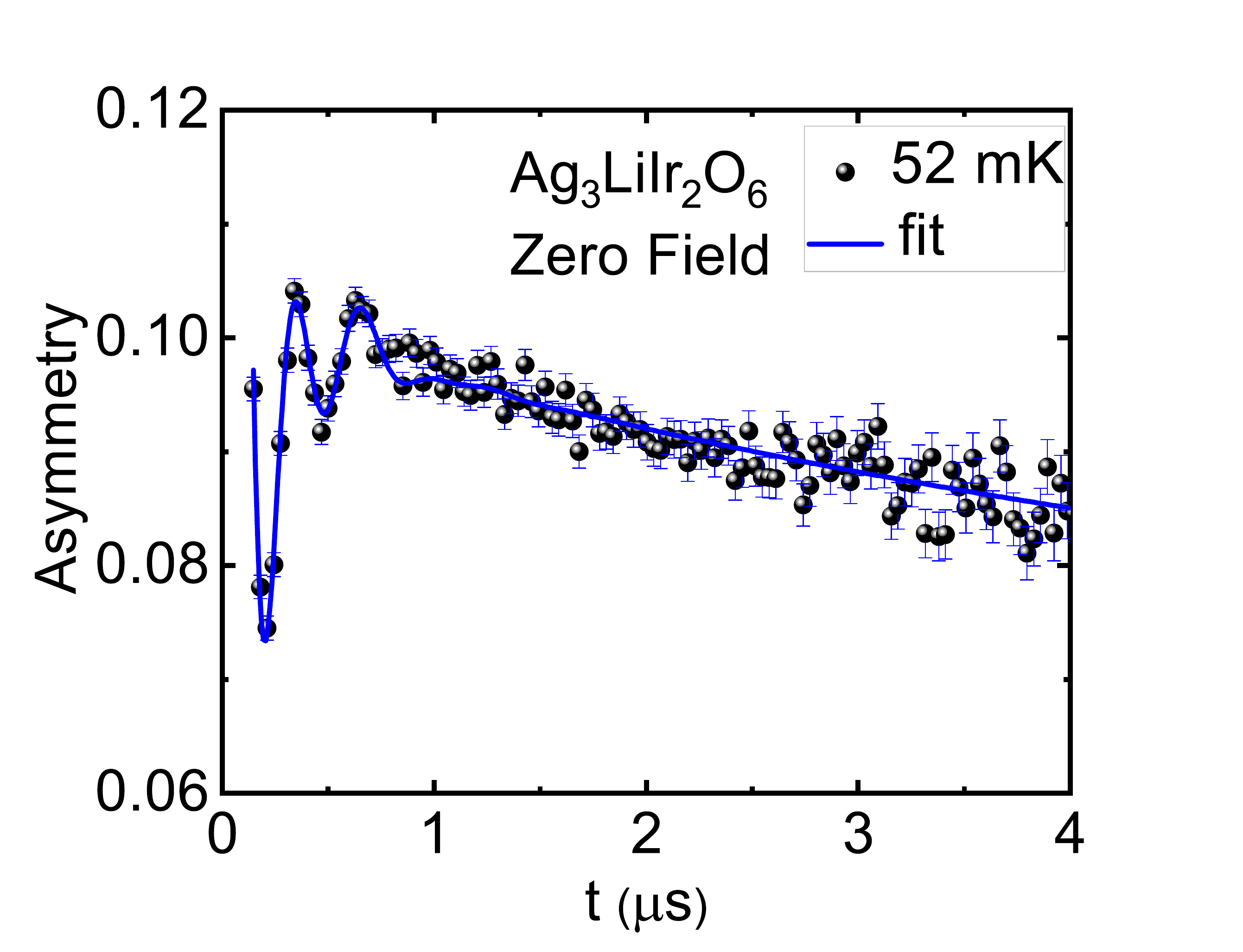}
	\caption{Variation of the muon asymmetry with time is shown at 52 mK 
together with a fit to Eq.~\ref{bessel}.  
The fit parameters are nearly unchanged up to 7 K.} 
	\label{musr-bessel}
\end{figure}

\begin{figure}[b]
	\includegraphics[width=1.0\columnwidth,clip=true,trim=0 0 30mm 8mm]{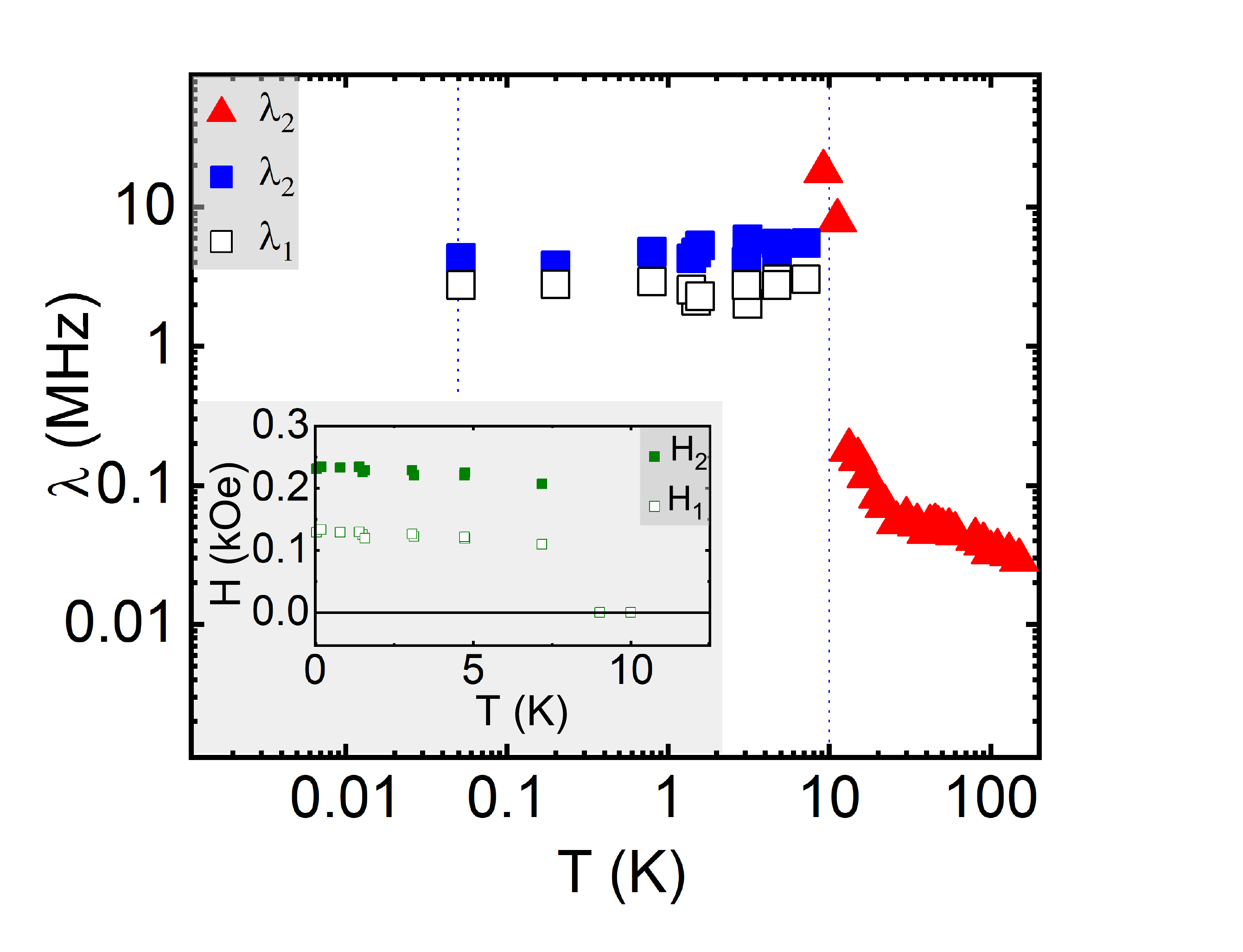}
	\caption{Variation of the muon relaxation rate (main figure) and the local field at the muon site (inset) as a function of temperature for ALIO is shown.
}
	\label{musr-lambda}
\end{figure}

The intermediate region of 13 K to 9 K shows a sharply falling 
muon asymmetry with $\lambda _2(T)$ showing a sharp increase 
with decreasing temperature as an approach to long-range order.  
Going further down in temperature, we find that below 9 K, there 
are clear oscillations in the muon asymmetry as a function of 
time.  This is a classic signature of the presence of 
long-range magnetic order.  The temporal decay of the muon 
asymmetry is nearly unchanged from 7 K down to 52 mK.  
The data in the range 52 mK-7 K were well 
fit (see Fig.~\ref{musr-bessel}) by the following equation:
\begin{equation}\label{bessel}
	\begin{split}
		A(t) & =  A_0 + A_1e^{-\lambda _1(T)t}J_0(\gamma H_1t) \\ 
		& +  A_2e^{-\lambda _2(T)t}J_0(\gamma H_2t)  +  A_3e^{-\lambda _3(T)t}  \\
	\end{split}
\end{equation}
where $\gamma$ is the muon gyromagnetic ratio ($\gamma =2\pi \times 135.539$ MHz/Tesla).

%%%%%%%%%%%%%%%%%%%%%%%%%%%%%%%%%%%%%%%%%%%%%%%%%%%%%%%%%%%%%%%%%%%%%%
%%%%%%%%%%%%%%%%%%%%%%%%%%%%%%%%%%%%%%%%%%%%%%%%%%%%%%%%%%%%%%%%%%%%%%%%%%%%%%%%%%%%%%%%%%%%
%%%%%%%%%%%%%%%%%%%%%%%%%%%%%%%%%%%%%%%%%%%%%%%%%%%%%%%%%%%%%%%%%%%%%%
%%%%%%%%%%%%%%%%%%%%%%%%%%%%%%%%%%%%%%%%%%%%%%%%%%%%%%%%%%%%%%%%%%%%%%%%%%%%%%%%%%%%%%%%%%%%
The $A_3e^{-\lambda _3(T)t}$ term is ascribed to muons which are initially parallel 
to the internal field components and hence do not precess. As for the other 
significant terms, $J_0(\gamma H_1t)$ and $J_0(\gamma H_2t)$ are zeroth order 
Bessel functions.  In case of ordering that is commensurate with the lattice, one 
expects an  exponentially damped sinusoidal variation of $A(t)$ in the ordered state. 
The Bessel function variation observed here is indicative of magnetic order 
incommensurate with the lattice~\cite{R.C.Williams2016} (such as for a spin density wave), where the muon 
experiences fields upto a maximum of $H_1$ or $H_2$, in the present case.  We found 
that the fit was better with two Bessel functions rather than one (see 
SM~\cite{SI} for a comparison), suggesting the 
presence of two types of magnetic environments for the muons.  These could arise 
either from the presence of two kinds of regions with different spin order, or 
possibly from crystallographically inequivalent muon stopping sites.  

From our DFT 
calculations (see Sec.~\ref{sec:DFT} 
%and Ref.~\cite{SI} 
for details), we conclude that while the 
N\'{e}el type order has the lowest energy, the stripy phase is not much higher in 
energy, which lends credence to the first possibility.  
Thus assuming this scenario of a single muon stopping site,
%two different magnetic environments seen by the muon, 
the site was determined from 
calculations of the electrostatic energy and was found to be about 1 \AA \,  
from the oxygen ion (see SM~\cite{SI}  for details).  This is similar 
to that found in cuprates and other oxide materials~\cite{Mohd_Tajudin_2014}.
We then calculated the  dipolar magnetic fields at the muon stopping site in N\'{e}el 
and stripy environments respectively.  The calculated field values of 139 Oe for 
the N\'{e}el phase and 266 Oe for the stripy phase (assuming a moment 
of 0.5 $\mu_B$/Ir which is typical for Ir$^{4+}$) are in reasonable register
with the values of 129 Oe and 232 Oe obtained as the 
averages of fit parameters  $H_1$ and $H_2$
respectively between 52 mK to 1.4 K (see inset of Fig.~\ref{musr-lambda}). 

Finally, we look at the variation of the muon relaxation rates
$\lambda _1$ and $\lambda _2$ vs $T$. It shows a peak at about 9 K, but does 
not fall to low values even at 52 mK as is expected to happen for progressively
slower dynamics as we go deeper into the ordered state (Fig.~\ref{musr-lambda}).  
It rather stays almost flat and undiminshed at a value of about 5 MHz in the
low temperature side as seen in Fig.~\ref{musr-lambda}.
We speculate that persistent spatio-temporal fluctuations of the 
stripy and N\'{e}el regions
are responsible for this. In such a scenario, this would present an 
interesting example where these fluctuations persist
even at temperatures more than two orders of magnitude lower than the 
transition temperature ($\sim$ 9 K) 
till at least a thermal energy scale of $\sim$ 50 mK (4.3 $\mu$eV). 
Could these then be quantum mechanical in origin?

From a quantitative point of view, the rather large value of 5 MHz for
the muon depolarisation rate is quite remarkable,
comparable to (or even larger than) those
seen in a variety of spin liquids (for example, see the review Ref.~\cite{ZhuZiHao2020} 
and the references therein; also see recent Refs.~\cite{Kundu2020, Kundu2020a} not 
covered in this review). Even in $\alpha$-RuCl$_3$ which has been established
to be magnetically ordered, the muon relaxation rate is appreciable 
($\sim 0.5$ MHz in a single crystal~\cite{muon_rucl} 
and $\sim 2$ MHz/$\sim 4$ MHz in a polycrystalline sample~\cite{Lang2016})
though smaller compared to ALIO. However, these
observations on $\alpha$-RuCl$_3$ are limited respectively 
to roughly $1/2$ the ordering temperature 
in the single crystal work~\cite{muon_rucl}, and about $1/5^{\text{th}}$ the 
ordering temperature in the polycrystalline sample study~\cite{Lang2016}. 
Whereas
our observations on ALIO go down to $1/200^{\text{th}}$ of the ordering temperature
and can be considered well-representative of the ground state physics.
This suggests that the persistent dynamics are really a feature of
the magnetically ordered many-body ground state.

\subsection{NMR}
\label{subsec:NMR}
%%%%%%%%%%%%%%%%%%%%%%%%%%%%%%%%%%%%%%%%%%%%%%%%%%%%%%%%%%%%%%%%%%%%%% 
\begin{figure}[t]
	\includegraphics[width=1\columnwidth,clip=true,trim=10mm 0 30mm 22mm]{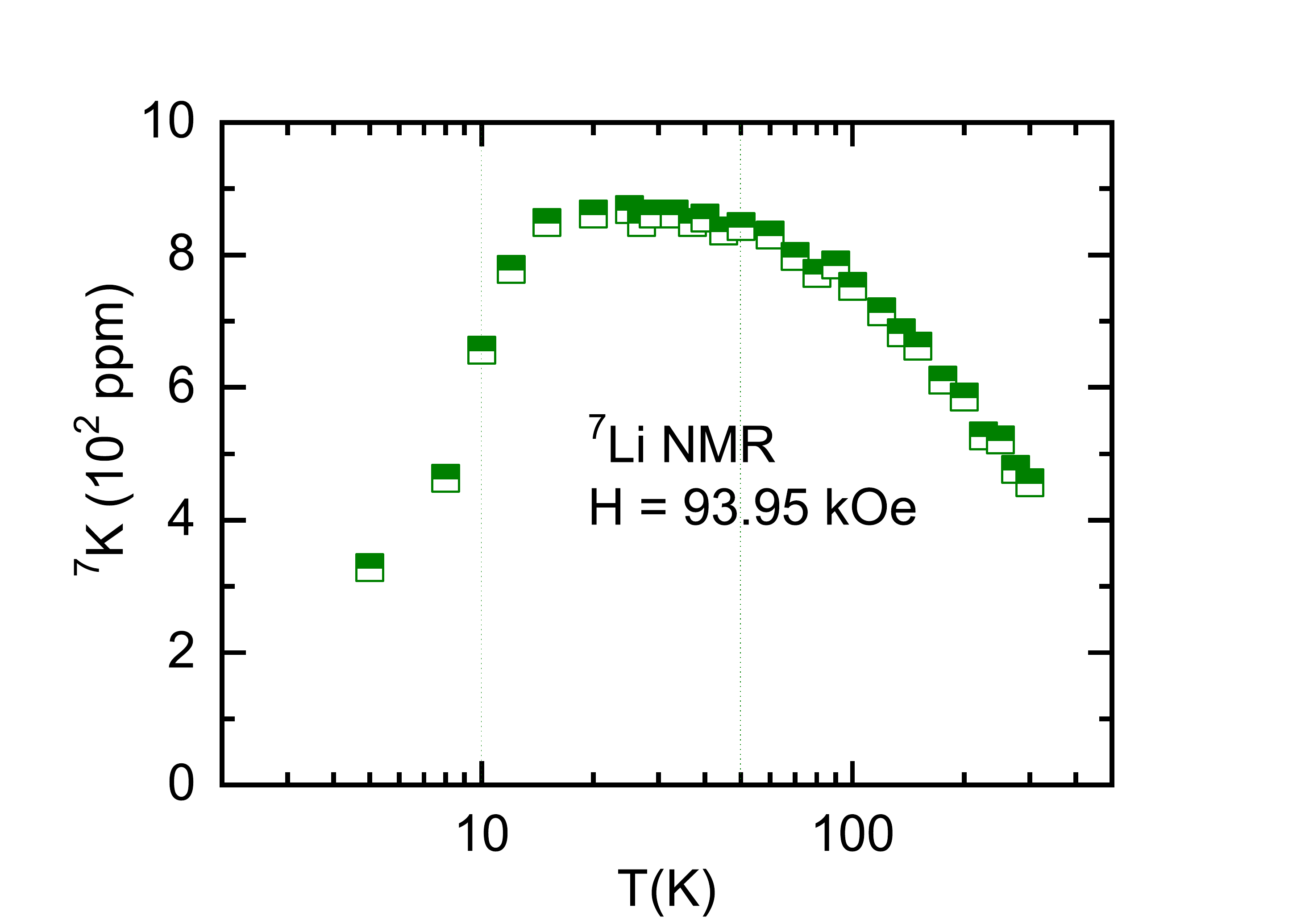}
	\caption{The $^7$Li NMR shift increases with decreasing temperature and then shows a broad plateau below 50 K.
%\SP{may be vertical lines showing 10 K and 50 K.}
}
	\label{nmrshift}
\end{figure}
\begin{figure}[b]
	\includegraphics[width=1.0\columnwidth,clip=true,trim=10mm 0 30mm 20mm]{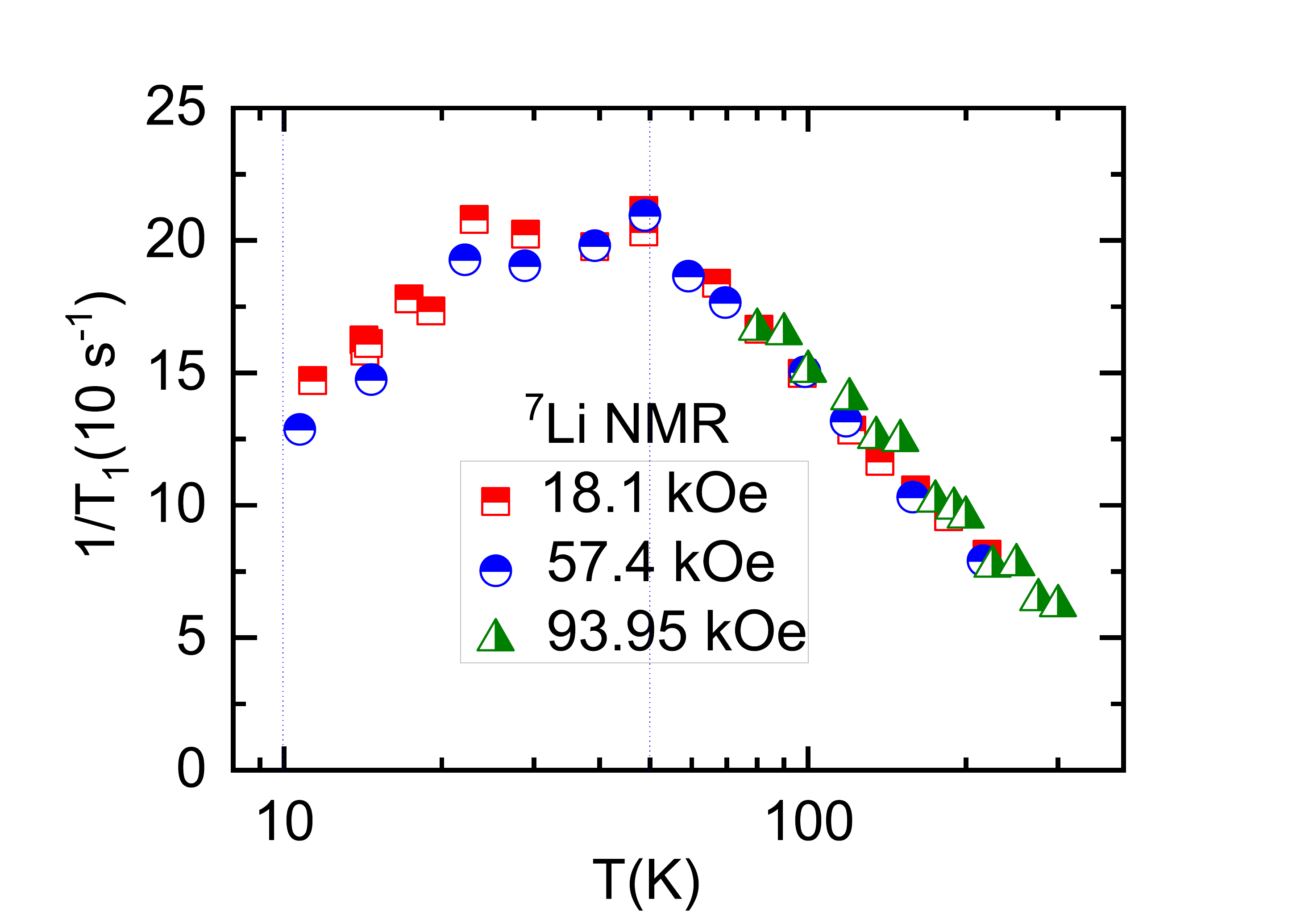}
	\caption{The variation of he $^7$Li NMR spin-lattice relaxation rate is shown with $T$.  This also shows a peak at about 50 K where there is a susceptibility anomaly.
%\SP{maybe the main panel is a zoomed-in version up to 70 K or around. and an inset
%for the whole range.}
}
	\label{spinlattice}
\end{figure}
%%%%%%%%%%%%%%%%%%%%%%%%%%%%%%%%%%%%%%%%%%%%%%%%%%%%%%%%%%%%%%%%%%%%%%%%%%%%%%%%%%%%%%%%%%%%%%%%%%%%%%%%%%%%%%%%%%%%%%%%%%%%%%%%%%%%%%%%%%%%

Having established the presence of magnetic order in ALIO from our zero 
field $\mu$SR data at low temperatures, we now move over to the  local probe 
technique of NMR to examine the variation of the intrinsic spin susceptibility 
in the paramagnetic state, as well as to look for complementary evidence of ordering. 
The bulk susceptibility on the other hand can have a low-$T$ upturn 
arising from extrinsic contributions or orphan spins which may not be reflective 
of the intrinsic properties.  We have therefore performed $^7$Li local probe 
NMR measurements to determine the shift of the $^7$Li resonance ($^7$K) with respect 
to a diamagnetic reference as a function of temperature. The results are shown 
in Fig.~\ref{nmrshift} where it is seen that the intrinsic susceptibility (in 
the form of a $^7$K line shift) increases with decreasing temperature and then 
exhibits a broad plateau region below about 50 K.  

Furthermore, as NMR is a good probe 
of low-energy excitations, we have performed $^7$Li NMR spin-lattice relaxation 
rate measurements as a function of temperature.  
The recovery of the longitudinal nuclear magnetisation after a saturating pulse sequence 
was well fitted with a single exponential.    As shown in Fig.~\ref{spinlattice}, we find 
that 1/$T_1$ increases with decreasing temperature  and has a peak around 50 K 
similar to systems which show ordering.  The $^7$K and 1/$T_1$ results are found 
independent of the applied field between 18-94 kOe.

We now contrast our observations with those reported 
in a very recent Ref.~\cite{wang2020} performed on 
two different batches of samples (A and B) of ALIO.  
Wang \textit{et al.} found that the cleaner sample A 
showed a single peak in the NMR spectrum as opposed to 
two peaks for sample B. The longitudinal nuclear magnetisation 
recovery for sample A was found to be single exponential 
in contrast to a stretched exponential variation for 
sample B. Also, the values of the relaxation rate were 
higher for sample A compared to sample B. The near absence 
of a second peak in the NMR lineshape of our sample of 
ALIO, together with a single exponential recovery in $T_1$ 
and with the absolute value of the $^7$Li NMR relaxation 
rate 1/$T_1$ on the higher side (similar to sample A of  
Ref.~\cite{wang2020}) suggests that our sample is of 
high quality.

%%%%%%%%%%%%%%%%%%%%%%%%%%%%%%%%%%%%%%%%%%%%%%%%%%%%%%%%%%%%%%%%%%%%%%%%%%%%%%%%%%%%%%%%%%%%%%%%%%%%%%%%%%%%%%%%%%%%%%%%%%%%%%%%%%%%%%%%%%%%
\begin{figure}[t]
	\includegraphics[width=1.0\columnwidth,clip=true,trim=0 0 0 15mm]{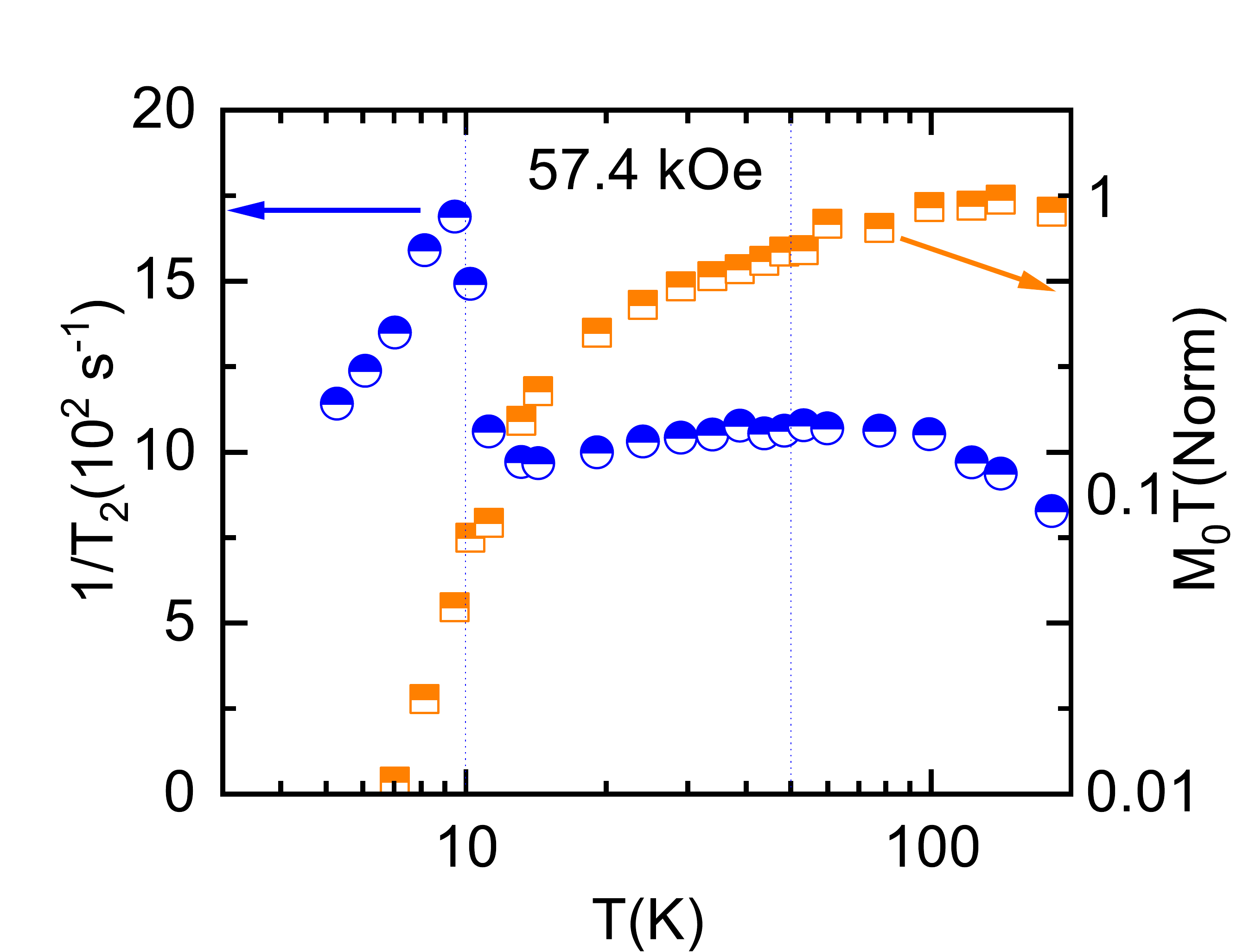}
	\caption{The variation of the $^7$Li NMR spin-spin relaxation rate (left $y$-axis) is shown with $T$.  The right $y$-axis shows the product of the nuclear magnetisation and temperature normalized to the maximum, as a function of $T$.}
	\label{spinspin}
\end{figure}
%%%%%%%%%%%%%%%%%%%%%%%%%%%%%%%%%%%%%%%%%%%%%%%%%%%%%%%%%%%%%%%%%%%%%% 

We also monitored the total NMR spectral intensity as a 
function of temperature which naturally involves the measurement of  the spin-spin 
relaxation rate (1/$T_2$).  The total spectral intensity should normally vary 
as a Curie law due to a Curie variation of the nuclear magnetisation.   
Hence, the product of the nuclear magnetisation $M_0$ and $T$ should remain 
constant with temperature in case the same number of nuclei contribute to the 
signal at all temperatures. Our results for 1/$T_2$ and $M_0T$ are shown in 
Fig.~\ref{spinspin}.  We observe an onset of a decrease in intensity already 
around 50 K and a near complete wipe-out below 10 K.    The observed wipe-out 
is a classic signature of the onset of 
long-range magnetic ordering.
An anomaly is also seen in the 1/$T_2$ data around 10 K. We thus conclude that there is 
an onset of short-range correlations around 50 K and eventually 
long-range magnetic ordering
sets in at about 10 K.  

\section{ First Principles Electronic Structure Calculations }
\label{sec:DFT}

In order to gain a microscopic understanding of the electronic and magnetic behavior, 
we first optimised the crystal structure parameters
using  first-principles density functional theory calculations.
The crystal structure of ALIO has been shown in Fig.~\ref{strucFig}, and 
we recall that the various exchange parameters, in particular the Kitaev exchange interaction 
are strongly dependent on the nearest-neighbor ($nn$) bond-length and bond-angles. 
To cross-check the experimental refinement of the
position of the light atoms (such as O) based on X-ray diffraction, 
we have independently determined the structural parameters by carrying out  
an ionic relaxation simulation for ALIO 
while maintaining the crystal symmetries of space group C2/m 
(see SM~\cite{SI} for details). 
A comparison of the structural data for the experimental and the relaxed structures 
are presented in Table~\ref{StructureRelax} along with the
structural data of H$_{3}$LiIr$_{2}$O$_{6}$ (HLIO) 
and the parent compound $\alpha$-LIO with similar stoichiometries.
We find for the relaxed structure of ALIO, 
the Ag-O bond lengths along the O-Ag-O are 
nearly equal and substantially larger in comparison to its H counterpart  
while the Ir-O-Ir angle for the Z-bond are nearly identical for both the systems. 
All the subsequent calculations are performed with this relaxed structure of ALIO.
\begin{table}[b]
	\caption{Bond lengths are given in \AA~and bond-angles are in degrees$(^\circ)$. }
	\centering
	\scalebox{0.83}{
		\begin{tabular}{|l |c | c | c | c|}
			\hline
			Parameters & $\alpha$-Li$_2$IrO$_3$ & H$_3$LiIr$_2$O$_6$ & Ag$_3$LiIr$_2$O$_6 $ & Ag$_3$LiIr$_2$O$_6 $ \\
			& (exp)~\cite{alphaLIO_structure}  & (exp)~\cite{Li_R_2018} & (exp)~\cite{Bette_C_2019} & (relax) \\
			\hline
			Ir-Ir distance & & & & \\ (X/Y-bond) & 2.98  &3.10 & 3.06 & 3.04\\
            \hline
			Ir-Ir distance & & & & \\ (Z-bond) & 2.98 & 3.05 & 3.04 & 3.08 \\
			\hline 
			Ir-O-Ir angle & & & & \\ (X/Y-bond) & 94.74  & 99.77 & 98.55 & 97.76  \\
            \hline
			Ir-O-Ir angle & & & & \\ (Z-bond) & 95.42 & 99.03 & 92.43 & 100.08 \\
			\hline
			(Li,H,Ag)-O & & & & \\ distance &  1.88, 2.13  &1.23, 1.27 & 1.94, 2.09 & 2.08, 2.10 \\
			\hline
		\end{tabular}
	}
	\label{StructureRelax}
	
\end{table}

\begin{figure}[t]
	\includegraphics[width=1\columnwidth]{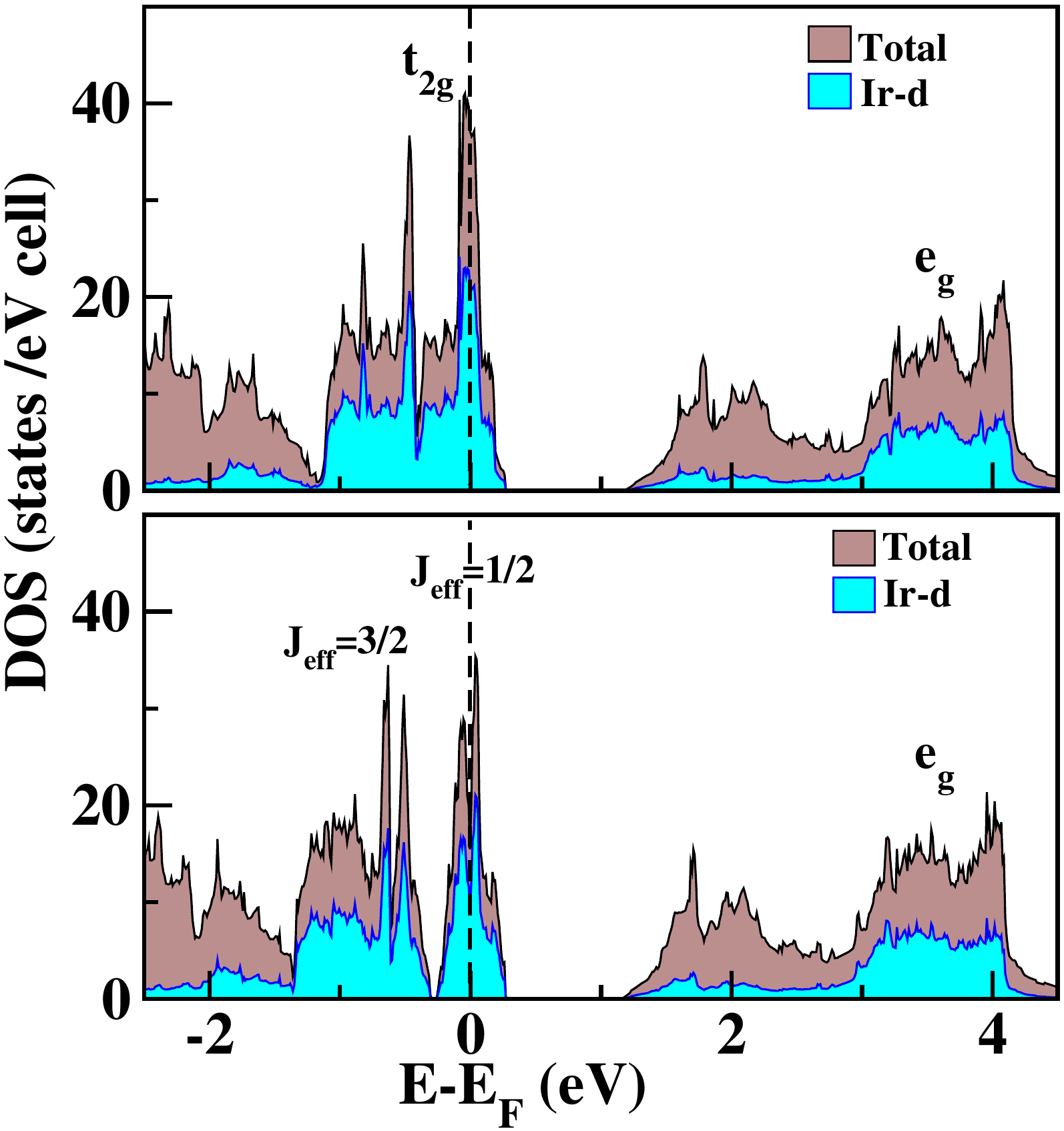}
	\caption{Non-magnetic total density of states (DOS) (brown shaded region) 
and partial Ir-$d$ DOS (cyan shaded region) are shown. 
Top and bottom panels show DOS for GGA calculations without and with spin-orbit coupling. }
	\label{irsoFig}
\end{figure}

We begin with an investigation of the electronic structure of ALIO 
without magnetic order. 
The results of our calculations are summarized in the
top panel of Fig.~\ref{irsoFig}, 
where we have plotted the total as well as the Ir projected density of states (DOS). 
We find %from Fig.~\ref{irsoFig} 
that the octahedral environment of Ir splits 
its $d$ states into $t_{\mathrm{2g}}$ and $e_{\mathrm{g}}$ states by a large 
crystal field splitting ($\Delta_{CF}$ $\sim$ 4.3~eV) characteristic of 
iridates. The $t_{\mathrm{2g}}$ states are further split  
due to monoclinic distortion and host the Fermi level. The oxygen states 
are completely occupied, while the Ir-$e_{\mathrm{g}}$, Li-$s$ and Ag-$d$ states 
are completely empty consistent with the nominal ionic formula 
Ag$^{1+}_{3}$Li$^{+1}$Ir$^{4+}_{2}$O$^{-2}_{6}$.

In view of the large $\Delta_{CF}$, the entire physics of these systems is 
essentially governed by the $t_{\mathrm{2g}}$ states. We have therefore 
constructed a low energy tight-binding model retaining only the $t_{\mathrm{2g}}$ 
states in a local basis (where the local $x$, $y$, and $z$ axes point towards 
the ligands) and downfolded all other higher degrees of freedom using the 
NMTO downfolding method~\cite{Andersen_M_2000}. The C2/m group splits 
the $t_{\mathrm{2g}}$ states retaining two-fold symmetry at each metal site. 
The various hopping interactions between the Ir atoms reveal that the 
hopping corresponding to the first $nn$ is stronger compared to the other interactions. 
The C2/m space group provides two types of symmetry inequivalent 
nearest-neighbor bonds: (i) Z bonds, parallel to the crystallographic $b$-axis, 
are of local C2h symmetry; (ii) X or Y bonds with lower symmetry. 
The nearest neighbor $d$-$d$ hopping integrals for the Z-bond expected for 
the ideal structure are extracted from NMTO downfolding calculations 
by suitable averaging and the values are  
$t_{1} \equiv t_{xz,xz} = t_{yz,yz}$ = -49.1 meV, 
$t_{2} \equiv t_{xz,yz} = t_{yz,xz}$ = 204.0 meV, 
$t_{3} \equiv t_{xy,xy}$ =-1.4 meV, 
$t_{4} \equiv t_{xz,xy} = t_{xy,xz} = t_{yz,xy} = t_{xy,yz}$ = 46.2 meV. 
While $t_{2}$ is strongest as expected, $t_{3}$ is however strongly suppressed. 
This will have a profound impact on the exchange intercations to be 
discussed later in the section. 

%%%%%%%%%%%%%%%%%%%%%%%%%
%%%%%%%%%%%%%%%%%%%%%%%%%
We now present the effect of spin-orbit coupling (SOC) on the electronic structure of ALIO.
The total and the Ir projected DOS are displayed in Fig.~\ref{irsoFig}. 
We find that the SOC further splits the $t_{2g}$ states into low-lying, 
four-fold degenerate $J_{\text{eff}}= 3/2$ and high-lying two-fold degenerate 
$J_{\text{eff}}= 1/2$ states. The five electrons of 
nominal Ir$^{4+}$ completely fill the $J_{\text{eff}}= 3/2$ states, while 
the $J_{\text{eff}}= 1/2$ state is half-filled which upon inclusion of a moderate 
Hubbard interaction $U$ makes the system insulating. 
In the limit $U >> t$, the holes occupying the $J_{\text{eff}}= 1/2$ states 
are nearly localized on the metal sites and the low energy degrees of 
freedom are pseudo-spin-1/2 variables {$\mathbf{S_{i}}$} connected to 
the $J_{\text{eff}}= 1/2$ states. 

This spin-orbit entangled pseudo-spin state of Ir atom on a honeycomb lattice 
in the strong coupling limit hosts bond dependent anisotropic Kitaev exchanges 
in addition to the usual isotropic Heisenberg exchange terms. The nearest 
neighbor spin Hamiltonian may be written as 
$H_{\text{spin}} = \sum_{\langle i,j \rangle} \mathbf{S}_{i} 
\cdot \mathbf{J}_{ij} \cdot \mathbf{S}_{j}$, 
where $\mathbf{J}_{ij}$ is a 3 $\times$ 3 symmetric matrix due to the 
presence of local inversion symmetry and is given by: 
\begin{equation}
		\label{bond_eqn}
	\begin{bmatrix} 
		J_{1} & \Gamma_{1} & \Gamma^{\prime}_{1} \\
		\Gamma_{1} & J_{1} & \Gamma^{\prime}_{1}\\
		\Gamma^{\prime}_{1} & \Gamma^{\prime}_{1} & J_{1}+K_{1} \\
	\end{bmatrix}
	\quad
\end{equation}

The various parameters of the spin Hamiltonian (Eq.~\ref{bond_eqn}) are 
calculated following Refs.~\cite{Winter_C_2016, Rau_G_2014} 
using the hoppings obtained from the NMTO downfolding method mentioned earlier
and neglecting the crystal field terms. 
For $U$ =1.7~eV, $J_{H}$ = 0.3~eV and $\lambda$ = 0.4~eV suitable for 
ALIO~\cite{Winter_C_2016},
we obtain for the spin Hamiltonian 
($J_{1}$, $K_{1}$, $\Gamma_{1}$ ,$\Gamma^{\prime}_{1}$) in Eq.~\ref{bond_eqn} 
as (+3.19, -11.4, -1.3, -2.99) meV and (2.86, -5.85, -0.67, -1.54) meV 
by using the strong coupling expressions 
from Ref.~\cite{Winter_C_2016} and Ref.~\cite{Rau_G_2014} respectively. 
The reported values of ($J_{1}$, $K_{1}$, $\Gamma_{1}$,$\Gamma^{\prime}_{1}$)
from two different studies on similarly stoichiometric HLIO 
structure that has been suggested to be a Kitaev quantum spin-liquid~\cite{Kitagawa_A_2018} 
are ($1.8$ meV, $-12.0$ meV, $-0.2$ meV, -3.2meV)~\cite{Yadav_S_2018} 
and ($-1.3$ meV, $-15.4$ meV, $+1.5$ meV, $-5.1$ meV)~\cite{Li_R_2018} 
respectively. While the ferromagnetic nature of $K_{\mathrm{z}}$ obtained for ALIO is similar 
to that of HLIO, the magnitude of $\vert \frac{K_{1}}{J_{1}} \vert$ is 2-3.5. This is much smaller in comparison to HLIO (8.5-12.5). 
It is therefore likelier that ALIO will  
order like the parent compound %$\alpha$-Li$_2$IrO$_3$.
$\alpha$-LIO~\cite{Williams_I_2016},
for which the (two) reported estimates of 
($J_{1}$, $K_{1}$, $\Gamma_{1}$, $\Gamma^{\prime}_{1}$) 
are ($-4.6$ meV, $-4.2$ meV, $+11.6$ meV, $-4.3$ meV) and
($-3.1$ meV, $-6.3$ meV, $+9.4$ meV, $-0.1$ meV)~\cite{Winter_C_2016}.

Perturbative analysis~\cite{Rau_G_2014} estimates that 
$J \sim \frac{t_{dd}^{2}}{U}$ and 
$K \sim \frac{t_{pd}^{4}}{\Delta_{pd}^{2}} \frac{J_{H}}{U^{2}}$ 
with the Heisenberg term predominantly governed by direct exchange, 
while the Kitaev interaction is mostly due to superexchange processes along the Ir-O-Ir paths. 
Here, $t_{dd}$ and $t_{pd}$ stand for the hopping amplitudes between 
$d$ orbitals of neighboring Ir ions and between Ir-$d$ and O-$p$ states respectively. 
$\Delta_{pd}$ is the charge-transfer energy. 
Unlike Li-$s$ in honeycomb $\alpha$-LIO, the Ag-$d$ orbitals form strong covalent 
bonds with ligand O-$p$ orbitals 
(seen from the crystal orbital Hamiltonian population (COHP) plot in SM~\cite{SI}) 
resulting in strong $d$-$p$ mixing.  
This modifies the Ir-O-Ir superexchange interaction and enhances the Kitaev term ($K$).
In comparison to $\alpha$-LIO ($\Gamma_{1}$= 11.6 meV)~\cite{Winter_C_2016}, 
the increased Ir-O-Ir bond angle in ALIO suppresses $t_{3}$ 
and leads to the reduced $\Gamma_{1}$ similar to HLIO. 
We have also calculated the second and third neighbor
interaction strengths and these are found to be much weaker, 
($J_{\mathrm{2nd}}< 6\% \text{ of } J_{\mathrm{nn}}$; 
$J_{\mathrm{3rd}}< 1\% \text{ of } J_{\mathrm{nn}}$). 
Our calculated parameters place ALIO in the vicinity of
the striped phase and 
the Neel (AFM) phase in the phase diagram reported in 
Ref.~\cite{Rau_G_2014}. The 120$^{\circ}$ magnetic phase is, as well, in close 
proximity for these parameters.

%%%%%%%%%%%%%%%%%%%%%%%%%%%%%%%%%%%%%%%%%%%%%%%%%%%%%%%%%%%%%%%%%%%%%%
\begin{figure}[t]
	\includegraphics[width=1\columnwidth]{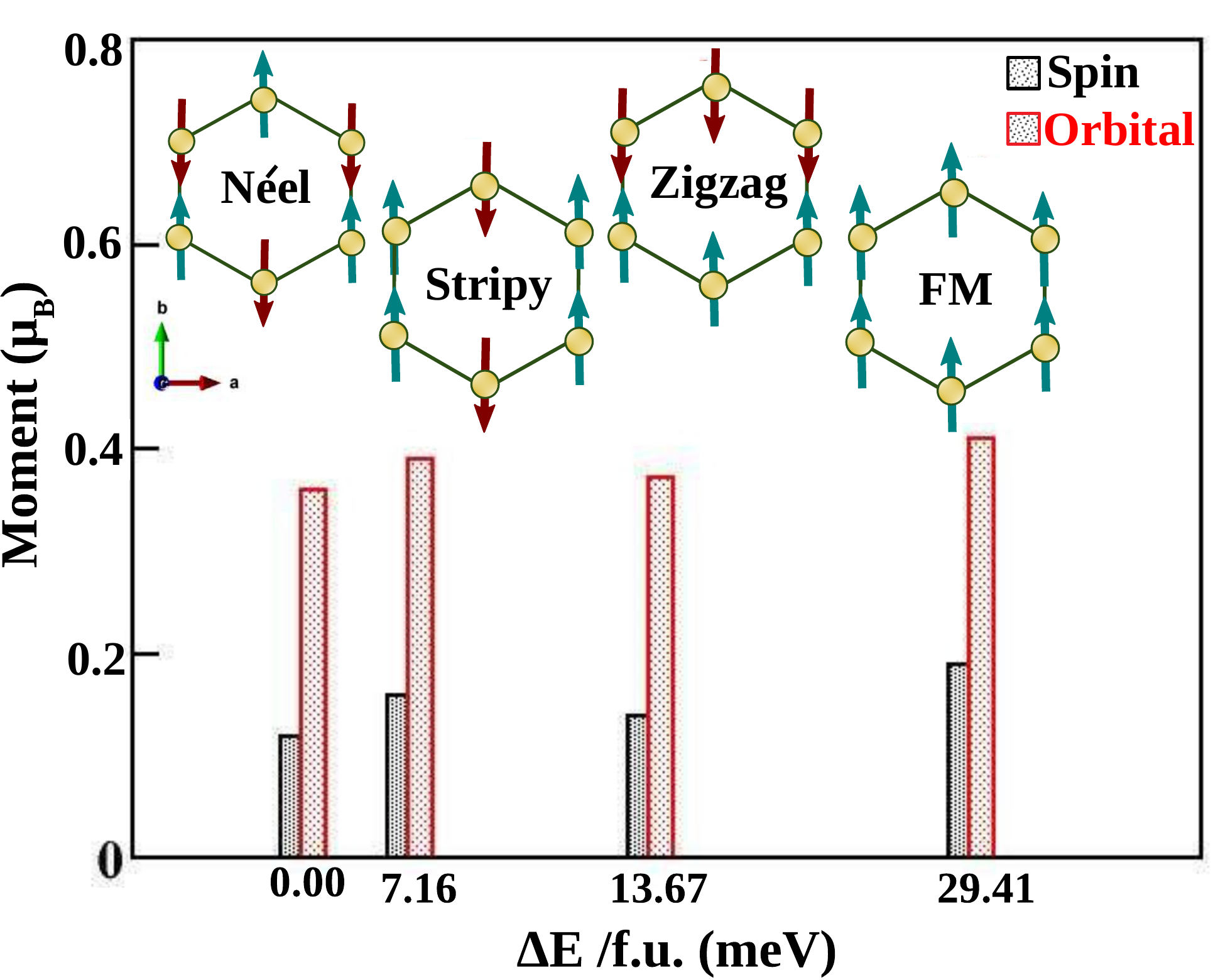}
	\caption{ Comparison of different magnetic configurations within GGA+SOC+U calculation
%\SP{fontsize improvement and inset arrows can be thicker.} \textit{Atasi: added}
}
	\label{magconfFig}
\end{figure}
%%%%%%%%%%%%%%%%%%%%%%%%%%%%%%%%%%%%%%%%%%%%%%%%%%%%%%%%%%%%%%%%%%%%%%
To determine the likely ground state of ALIO, 
we have considered several magnetic configurations 
whose spin and orbital moments are consistent with the
$J_{\text{eff}}=1/2$ state of Ir,
and calculated
their energies within GGA+SOC+U scheme (see Fig.~\ref{magconfFig}). 
Apart from the zigzag AFM order that has been observed in the parent compound
$\alpha$-LIO~\cite{Williams_I_2016}, we have also examined  three 
other representative magnetic 
orders: ferromagnet (FM), N\'{e}el AFM, and stripe AFM 
which have been observed in honeycomb materials for 
the $q=0$ magnetic structure~\cite{Hou_U_2017, Hou_F_2018}.
Whereas the N\'{e}el AFM order has the lowest energy among the 
chosen configurations, the stripy phase is not much higher 
in energy. The $q\ne0$ solutions and the incommensurate phases are 
not taken into consideration which may also dictate the ground state magnetism,
especially the 120$^{\circ}$ ordered phase.

%%%%%%%%%%%%%%%%%%%%%%%

\section{Discussion on Bulk Probes and high-temperature anomaly}
\label{sec:bulk_probes}

%%%%%%%%%%%%%%%%%%%%%%%%%%%%%%%%%%%%%%%%%%%%%%%%%%%%%%%%%%%%%%%%%%%%%%
\begin{figure}[b]
	\includegraphics[width=1\columnwidth]{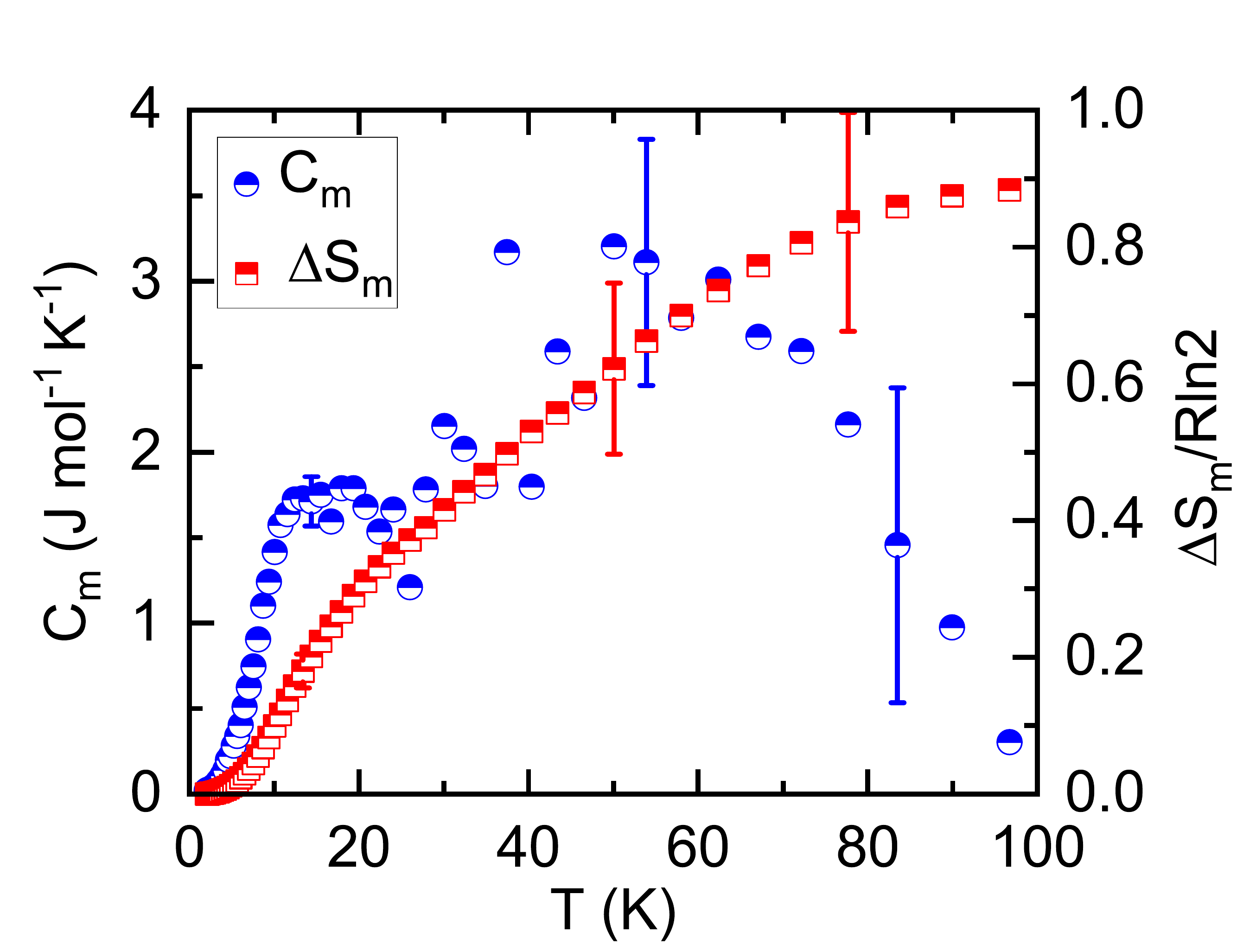}
	\caption{The magnetic heat capacity (C$_m$) and the entropy change (S$_m$) are 
		plotted in units of Rln2 as a function of temperature. Representative 
error bars are given at a few temperatures. Anomalies are 
observed at T$_H \simeq$ 50 K  and T$_L \simeq$ 13 K.}
	\label{HeatCapacity}
\end{figure}
\begin{figure}[t]
	\includegraphics[width=1\columnwidth]{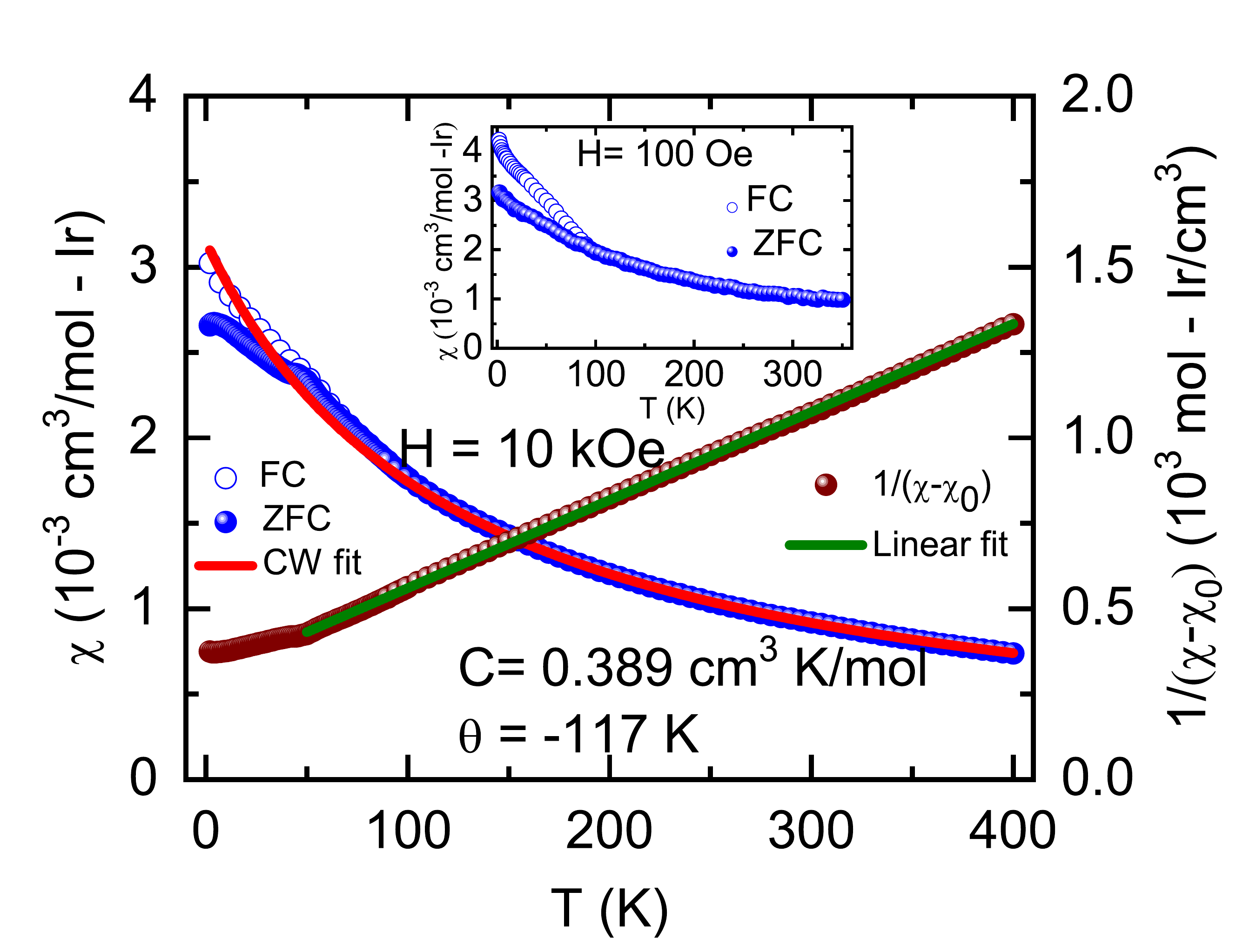}
	\caption{Variation of $\chi$ with temperature is shown together 
with the inverse susceptibility. The inset shows that the  susceptibility 
measured in a field of 100 Oe has a clear bifurcation below 100 K 
suggesting %disordered/glassy (static) magnetism.
static moment formation.}
	\label{zfcfc}
\end{figure}

%%%%%%%%%%%%%%%%%%%%%%%%%%%%%%%%%%%%%%%%%%%%%%%%%%%%%%%%%%%%%%%%%%%%%%%%%
In this section, we discuss bulk probes 
--- heat capacity (Fig.~\ref{HeatCapacity}) and magnetic 
susceptibility (Fig.~\ref{zfcfc}) ---,
and the high temperature anomaly in the NMR data in light of the low-temperature
ordered state established in the previous sections.
We start with heat capacity data. As is well-known, 
the measured heat capacity contains a contribution from the crystal lattice degrees
of freedom in addition to magnetic contributions.
At higher temperatures, it is the lattice constribution which generally
dominates the heat capacity.
To determine the lattice part, one can make use of structurally analogous 
non-magnetic variants of the given compound if available.  In such 
cases, corrections are necessary in inferring the lattice contribution 
of the magnetic compound from that of the non-magnetic analogue 
to account for the differences in their
effective Debye temperatures of the two compounds due to the different ionic masses
and unit cell volumes.   
Following such a correction, the heat capacities of the two compounds must 
coincide in a high-temperature region where the magnetic contribution is 
negligible.  In Ref.~\cite{Bahrami_T_2019,Bahrami2020}, 
the authors have made use of Ag$_3$LiSn$_2$O$_6$ (ALSO) as the nonmagnetic analog.  
However, it 
appears that no correction has been applied before subtracting these data 
from the measured heat capacity of ALIO.  This can be seen in Fig.~3(a) of 
Ref.~\cite{Bahrami_T_2019} and  Fig.~2(a) of Ref.~\cite{Bahrami2020} where 
the data for of ALIO and ALSO do not appear to overlap at high-temperatures. 

The magnetic specific heat of our sample, using the  data for the 
structurally analogous nonmagnetic  Ag$_3$LiTi$_2$O$_6$ (ALTO)  for the 
lattice contribution (see SM~\cite{SI} for details), is given in Fig.~\ref{HeatCapacity}.   
Anomalies are seen in our data at about 13 K and 50 K.  
Based on our NMR and $\mu$SR data described in Sec.~\ref{sec:NMR_muSR}, 
the lower-temperature anomaly is from long-range order. 
The higher-temperature 50 K anomaly could be from short-range correlations 
(recall the broad plateau seen in the $^7$Li NMR shift with temperature,
Fig.~\ref{nmrshift}). 
However, it should be noted that at high temperatures, 
the magnetic heat capacity is quite small compared 
to the lattice contribution, and hence a small error in the lattice heat 
capacity gives rise to a large error in the magnetic heat capacity.  
For instance, at about 50 K, the inferred magnetic specific heat is only 
about 13 \% of the total specific heat. Therefore, a 5 \% error in the lattice 
heat capacity will result in nearly 50 \% error in magnetic heat capacity at 50 K. 
Consequently, the magnetic entropy change as well suffers from high uncertainty, 
especially in the high-temperature region.  As a result, the inference of 
a two-stage entropy release is suspect in our opinion which calls
into question an interpretation in terms of localised and itinerant Majorana excitations.

%%%%%%%%%%%%%%%%%%%%%%%%%%%%%%%%%%%%%%%%%%%%%%%%%%%%%%%%%%%%%%%%%%%
%%%%%%%%%%%%%%%%%%%%%%%%%%%%%%%%%%%%%%%%%%%%%%%%%%%%%%%%%%%%%%%%%%%%%%%%%%%%%%%%
The DC magnetic susceptibility $\chi(T)=M/H$ of Ag$_3$LiIr$_2$O$_6$  
is shown in Fig.~\ref{zfcfc}.
It varies in a Curie-Weiss manner at high-temperature but has an anomaly at about 
50 K.  In particular, as shown in the inset of Fig.~\ref{zfcfc}, in a low field of 100 Oe,
there is a clear bifurcation between the zero-field-cooled ZFC and 
field-cooled FC curves a little below 100 K. 
Such a measurement is currently not available in the literature
for comparison.
%such a measurement was not reported in Ref.~\cite{Bahrami_T_2019}), 

Given the above bulk probe data, we revisit the high temperature anomaly 
seen in NMR data which one might want to optimistically interpret as having
to do with Kitaev physics.
Broad anomalies, such as the one in the $^7$Li NMR line shift of ALIO 
at about 50 K (Fig.~\ref{nmrshift}), are expected in low-dimensional quantum magnets with 
antiferromagnetic couplings at temperatures on the 
scale of the dominant exchange coupling.  In such a case, an anomaly 
is also expected in the temperature dependence of 1/$T_1$. It should be 
noted that $^7$Li nuclei are located symmetrically with respect to 
the Ir moments on the lattice, which will filter out N\'eel-type 
antiferromagnetic fluctuations.  However, given the evidence from 
$\mu$SR in Sec.~\ref{subsec:muSR} of the eventual incommensuration 
below 9 K, as also the possibility of the (incommensurate) stripy 
phase as a competing phase for which the fluctuations are not 
cancelled at the $^7$Li site, one expects now incomplete filtering 
at the $^7$Li site. The appearance of a peak at 50 K (Fig.~\ref{spinlattice})
then likely reflects a build-up of short-range dynamic magnetic correlations at the 
dominant exchange scale as concluded in Sec.~\ref{subsec:NMR}. 
This 50 K scale is indeed in consistent
with our DFT estimates (Sec.~\ref{sec:DFT}) for the various
exchange couplings (also, see SM~\cite{SI}).
On the other hand, as discussed in the NMR section (Sec.~\ref{subsec:NMR}), 
the absence of a sharp divergence in 1/$T_1$ at the ordering 
temperature ($\sim$ 10 K) is likely due to a wipe-out effect where a 
large fraction of the $^7$Li nuclei (nearly 90 \% as seen 
from Fig.~\ref{spinspin}) are already out of the window of observation by 10 K.

We now return to the bifurcation seen in the low-field ZFC/FC 
susceptibility data at about 100 K suggesting a freezing of 
magnetic moments (inset of Fig.~\ref{zfcfc}).  In higher fields, 
this anomaly moves to about 50 K. If this arose from intrinsic 
regions,  it would result in a sharp loss of the NMR spectral 
intensity at this temperature. On the other hand, the observed 
loss of NMR intensity is rather gradual as the temperature 
decreases from 100 K (Fig.~\ref{spinspin}). Further, one should 
have seen a reflection of the freezing of moments in 
the $\mu$SR relaxation rate $\lambda_2$ which is a 
zero-field measurement.  However, $\lambda_2$ shows a rather 
gentle variation  with temperature with a sharper increase 
only below 20 K (Fig.~\ref{musr-lambda}).  This then suggests 
that the ZFC/FC bifurcation is extrinsic in origin, possibly 
related to moments localized at stacking faults that naturally 
occur these systems, and one has to look for a different 
cause for the decrease in the NMR intensity.

We note that such a decrease of the NMR intensity has been 
seen in the heavy Fermion superconductor 
CeCu$_2$Ge$_2$ \emph{above} the superconducting transition 
temperature as well as in its non-superconducting 
variants~\cite{Nakamura_1992,Kitaoka_1991}. In these cases, 
it was ascribed to dynamical magnetic correlations arising 
from the itinerant physics of the ``heavy-fermion band magnetism". 
In the present case, our DFT estimates of the Heisenberg and 
Kitaev terms tell us that they are 
comparable ($\vert \frac{K_{1}}{J_{1}} \vert$ is about 2-3.5) 
and in the range of the energy scale $\sim$ 50 K. 
As discussed earlier in this section, these competing interactions
can be driving the development
of short-range dynamical magnetic correlations in ALIO. 
Whether there is some 
form of itinerancy -- by which, we have in mind some form 
of a quasi-particle excitation continuum~\cite{Kumar_2020} -- in 
the ALIO system driven by the Kitaev terms leading to the 
above decrease of the NMR intensity is an open question.

\section{Conclusions}
\label{sec:conclu}

In summary, ALIO is shown to exhibit magnetic long-range order below 9 K.  
This observation is consistent with our DFT calculations which find comparable 
Heisenberg and Kitaev exchange couplings.  Though $5d$ Ir$^{4+}$ possesses 
strong spin-orbit coupling that lie at the origin of substantial bond-dependent 
anisotropic Kitaev exchange terms, however they are tamed
by the Heisenberg exchange term in this system which leads to magnetic long-range order. 
More specifically, the  introduction of the heavy Ag atom in ALIO in place of 
Li in $\alpha-$Li$_2$IrO$_3$ or H in HLIO 
strongly affects the local structure and
enhances the interlayer Ag-$d$ and O-$p$ hybridization along linear O-Ag-O bonds.
%which is evident from our COHP calculation.
The increased bond angle originating due to Ag-O electronic repulsion
essentially contributes to the nearest-neighbor Kitaev exchange.
The nearest-neighbor Kitaev and Heisenberg exchanges are found to be ferromagnetic  
and antiferromagnetic respectively. Although a comparison of the energy 
among various magnetic configurations within the unit
cell finds the  N\'eel state  to be the lowest in energy, 
the stripy ordered state is not very far in energy (Fig.~\ref{magconfFig}).  
Both $J$ and $K$ parameters strongly depend on the
local geometry, an interesting aspect that can be explored in  
future studies by the tuning of bond-length and bond-angles driven by
pressure or the application of a magnetic field. 
This can open up the possibility of suppressing  $J$ to negligible values so 
that the Kitaev term $K$ dominates the physics of the ground state which 
might then be a Kitaev spin-liquid.

We end with the most unusual finding from our study: 
the continued presence of a high muon relaxation rate ($\sim 5$ MHz) %$\lambda$ 
deep into the ordered state as exemplified by Fig.~\ref{musr-lambda}.
Our observations go down to temperatures as low as 52 mK which is $1/200^{\text{th}}$ of
the ordering temperature as highlighted earlier in the introduction
and the text surrounding Fig.~\ref{musr-lambda}. 
This points to considerable dynamics of the ordered moments 
even in the ground state.
The $\alpha$-RuCl$_3$ system also shows a significant
muon depolarisation rate on the ordered side~\cite{muon_rucl,Lang2016}, however
as mentioned in Sec.~\ref{subsec:muSR}, these datasets are limited to
about $1/5^{\text{th}}$ of the ordering temperature.
A somewhat similar conclusion of persistent dynamics 
has been drawn~\cite{wang2020,imai2021_aps} 
based on an analysis of the $^7$Li NMR 1/$T_1$ in the ordered state of ALIO.
The natural question is what is the source of this dynamics.
Our data taken together with our computations of the muon stopping site and their dipolar fields 
are consistent with the co-existence of  N\'eel and stripe ordered domains. 
We conjecture that the persistent dynamics are due to
spatio-temporal 
fluctuations of these two kinds of ordered domains
that are likely driven by quantum effects in presence of non-neglible Kitaev terms.
This is a well-motivated question for theory, i.e. what is 
the effect of quantum fluctuations coming from the spin-flip Kitaev terms
in the ordered regions of the phase diagram~\cite{Rau_G_2014}.
Ultimately, an explanation for the persistent dynamics whether driven by Kitaev
terms or not is needed.

\section{acknowledgments}

We thank Nandini Trivedi and Vikram Tripathi for discussions. 
We thank MoE and Department of Science and Technology, Govt. of India
for financial support. 
 A.V.M. kindly acknowledges the Alexander von Humboldt foundation 
for the financial support for the renewed research stay at the University of Augsburg in 2018.
N.B. and A.V.M. kindly acknowledge support from the German 
Research Society (DFG) via TRR80 (Augsburg, Munich). 
Author group from IIT Bombay acknowledges support of measurement facilities at their institution.
Experiments at the ISIS Neutron and Muon Source
were supported by a beam-time allocation RB2068009 from the Science
and Technology Facilities Council~\cite{Mahajan2020}.
S.P. acknowledges financial
support from IRCC, IIT Bombay (17IRCCSG011) and
SERB, DST, India (SRG/2019/001419). 
I.D. thanks Department of Science and Technology, 
Technical Research Centre (DST-TRC) and Science and Engineering Research Board (SERB) India
(Project No. EMR/2016/005925) for support.

%{\bf References}
\bibliographystyle{apsrev}
\bibliography{ref}

\end{document}

%\begin{center}
%\Large {{\bf Supplementary Information}}
\title{\bf Supplemental Material: Unusual spin dynamics in the low-temperature magnetically ordered state of 
		Ag$_{3}$LiIr$_{2}$O$_{6}$}
	
	\author{Atasi Chakraborty} 
	\thanks{Equal contribution authors} 
	\affiliation{School of Physical Sciences, Indian Association for the Cultivation of Science, Jadavpur, Kolkata 700032, India}
	
	\author{Vinod Kumar} 
	\thanks{Equal contribution authors} 
	\affiliation{Department of Physics, Indian Institute of Technology Bombay, Powai, Mumbai 400076, India}
	
	\author{Sanjay Bachhar} 
	\affiliation{Department of Physics, Indian Institute of Technology Bombay, Powai, Mumbai 400076, India}
	
	\author{N. B\"{u}ttgen}
	\affiliation{Experimentalphysik V, Elektronische Korrelationen und Magnetismus, Institut f\"{u}r Physik, Universit\"{a}t Augsburg, 86135 Augsburg, Germany}
	
	\author{K. Yokoyama}
	\affiliation{ISIS Pulsed Neutron and Muon Source, STFC Rutherford Appleton Laboratory,	Harwell Campus, Didcot, Oxfordshire OX110QX, UK}
	
	\author{P. K.  Biswas}
	\affiliation{ISIS Pulsed Neutron and Muon Source, STFC Rutherford Appleton Laboratory,	Harwell Campus, Didcot, Oxfordshire OX110QX, UK}
	
	\author{V. Siruguri}
	\affiliation{UGC-DAE-Consortium for Scientific Research Mumbai Centre, Bhabha Atomic Research Centre, Mumbai 400085, India}
	
	\author{Sumiran Pujari}
	\affiliation{Department of Physics, Indian Institute of Technology Bombay, Powai, Mumbai 400076, India}
	
	\author{I. Dasgupta}
	\affiliation{School of Physical Sciences, Indian Association for the Cultivation of Science, Jadavpur, Kolkata 700032, India}

	\author{A.V. Mahajan}
	\thanks{corresponding author}
	\affiliation{Department of Physics, Indian Institute of Technology Bombay, Powai, Mumbai 400076, India}

	\date{\today}
%\end{center} 
\maketitle

\section{Sample Preparation}
The sample preparation of Ag$_3$LiIr$_2$O$_6$ (ALIO) has been done in two steps. The first step was to prepare the precursor $\alpha$-Li$_2$IrO$_3$ by a solid state reaction \cite{Dalton_2017}. After preparing phase pure $\alpha$-Li$_2$IrO$_3$, we mixed AgNO$_3$ in it (in a 1:3 molar ratio) and ground the mixture.  This was followed by heating at 200$^\circ$C for 6\,hrs. In the final step, we washed the product with deionized water several times to remove the excess nitrates and silver. The absence of nitrates was checked with a solution of KCl. 

The resulting product was  Ag$_3$LiIr$_2$O$_6$ (ALIO) as verified by lab x-ray diffraction measurements using a PANalytical X'Pert PRO diffractometer using Cu-K${\alpha}$ radiation ($\lambda=1.54182$\AA). Small amounts  of residual Ag and Li$_2$IrO$_3$ were detected in the x-ray diffraction pattern. Ag$_3$LiTi$_2$O$_6$ (ALTO) was prepared in a similar manner.

%We first prepared $\alpha$-Li$_2$IrO$_3$ as follows.  We took Li$_2$CO$_3$  and Ir-powder in stoichiometric amounts together with a 5\%  excess of Li$_2$CO$_3$ to compensate for any loss due to volatalization.  

%This was heated to 650$^\circ$C at the rate of 50$^\circ$C/hr and held there for 15\,hrs. The sample was then reground and heated to 1050$^\circ$C at the rate of 120$^\circ$C/hr and held there for 24\,hrs. After regrinding, this was followed by two rounds of  heating at 900$^\circ$C for 15\,hrs. 

\section{Rietveld Refinement}   

The  X-ray Diffraction data were taken at room temperature on a powder specimen. Table \ref{table7} summarizes the unit cell paramters and quality factors for the Rietveld refinement of Ag$_3$LiIr$_2$O$_6$. A 2D ordering peak, also known as the Warren peak is observed in the range 19\textdegree to 24\textdegree, shown in inset of Fig.\ref{XRDrefinement} with a fit \cite{Bahrami_T_2019} as given below,
\begin{equation}
	I_w(2\theta)= A e^{-g(2\theta)^2}+ B/(C+(2\theta)^2),	
\end{equation}

where A,B and C are constants, $ g $ is the exponent of the Gaussian term and it measures the
percentage of the stacking faults known as the $ g $-factor:
\begin{equation}
	g=\delta^2/d^2, 	\delta^2 = <d^2>-<d>^2
\end{equation}
where $ d $ is the interlayer spacing. A good fit with exponent, $ g=0.06(6) $ in the inset of Fig.\ref{XRDrefinement}  corresponds to at least 6\% volume fraction of
stacking disorder. After excluding the Warren peak, Rietveld refinemenet is performed with faultless model (stacking faults are ignored) to extract different quality factors, atomic coordinates, site occupancies, and the isotropic Debye-Waller factors (B$_{iso}$ = 8$\pi^2$U$_{iso}$) of Ag$_3$LiIr$_2$O$_6$,which are tabulated in Table \ref{table7} and Table \ref{table8} respectively.

\begin{table}[h!]
	\caption{Unit cell parameters and quality factors are reported for the Rietveld refinement of Ag$_3$LiIr$_2$O$_6$ at room temperature.}
	\centering
	\scalebox{1.0}{
		\begin{tabular}{|l c| c c|}
			\hline
			Unit Cell Parameters for space group C2/m& & Quality Factors&\\
			\hline
			a(\AA) & 5.277(7) &  &  \\
			\hline
			b(\AA)& 9.136(8)& R$_{Bragg}$ (\%)&5.61  \\
			\hline
			c(\AA)&6.493(8) & R$_F$ (\%) &3.82 \\
			\hline
			$\alpha$=$\gamma$  (\textdegree)&90 & R$_{exp}$ (\%)&5.8\\
			\hline
			$\beta$ (\textdegree)& 106.01(7) & R$_p$ (\%)&23.1\\
			\hline
			Z  & 2 & R$_{wp}$ (\%) & 19.9\\
			\hline
			V (\AA$^3$) & 300.98(0) & $\chi^2$ & 11.7\\
			\hline	
		\end{tabular}
	}
	\label{table7}
\end{table}

\begin{figure}[h!]
	\includegraphics[width=1.0\columnwidth]{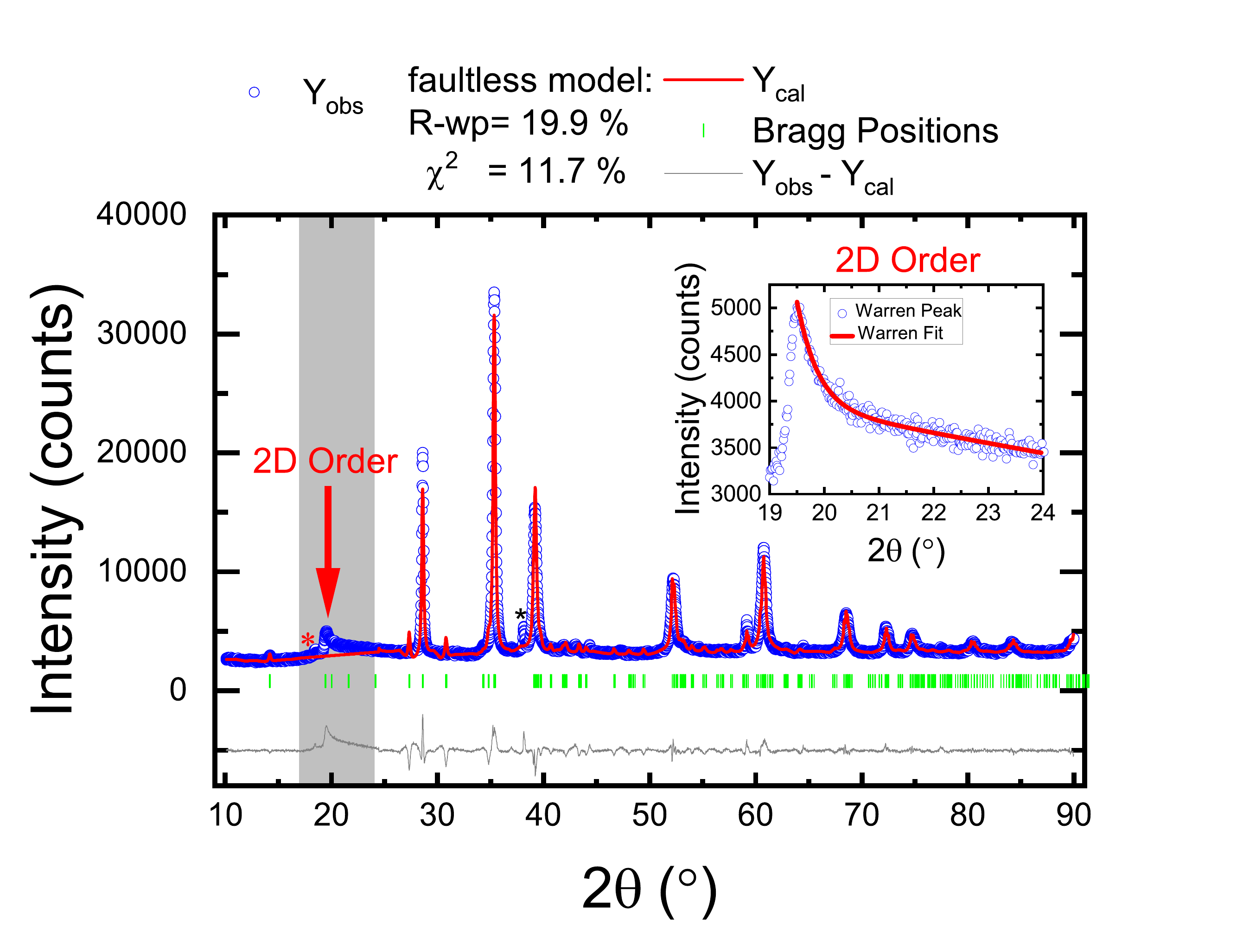}
	\caption{Rietveld refinement of the xrd pattern of Ag$_3$LiIr$_2$O$_6$ by using the
		ideal, faultless monoclinic structures (Space group: C2/m). \textcolor{red}{*},* indicates extrinsic peaks from residual $\alpha$-Li$_2$IrO$_3$ and Ag, respectively.}
	\label{XRDrefinement}
\end{figure}

\begin{table}[h!]
	\caption{Atomic coordinates, Normalized site occupancies, and the
		isotropic Debye-Waller factors (B$_{iso}$ = 8$\pi^2$U$_{iso}$)  are reported
		for the Rietveld refinement of Ag$_3$LiIr$_2$O$_6$.}
	\centering
	\scalebox{1.4}{
		\begin{tabular}{|l| c| c| c| c| c| c|c|}
			\hline
			Atom & Wyckoff position & Site &x&y&z&Norm. Site Occ.&B$_{iso}$(\AA$^2$)\\
			\hline
			Ir(1)&4g& 2&0&0.334&0&1&0.4\\
			\hline
			Li(1)&2a& 2/m&0&0&0&1&0.4\\
			\hline
			O(1)&4i&m&0.408&0&0.222&1&0.5\\
			\hline
			O(2)&8j&1&0.394&0.332&0.183&1&0.5\\
			\hline
			Ag(1)&4h&2&0&0.163&1/2&1&0.5\\
			\hline
			Ag(2)&2d&2/m&0&1/2&1/2&1&0.5\\
			
			\hline	
		\end{tabular}
	}
	\label{table8}
\end{table}
\section{Muon Spin Relaxation ({$\mu$}SR)}
\begin{figure}[h!]
	\includegraphics[width=1.0\columnwidth]{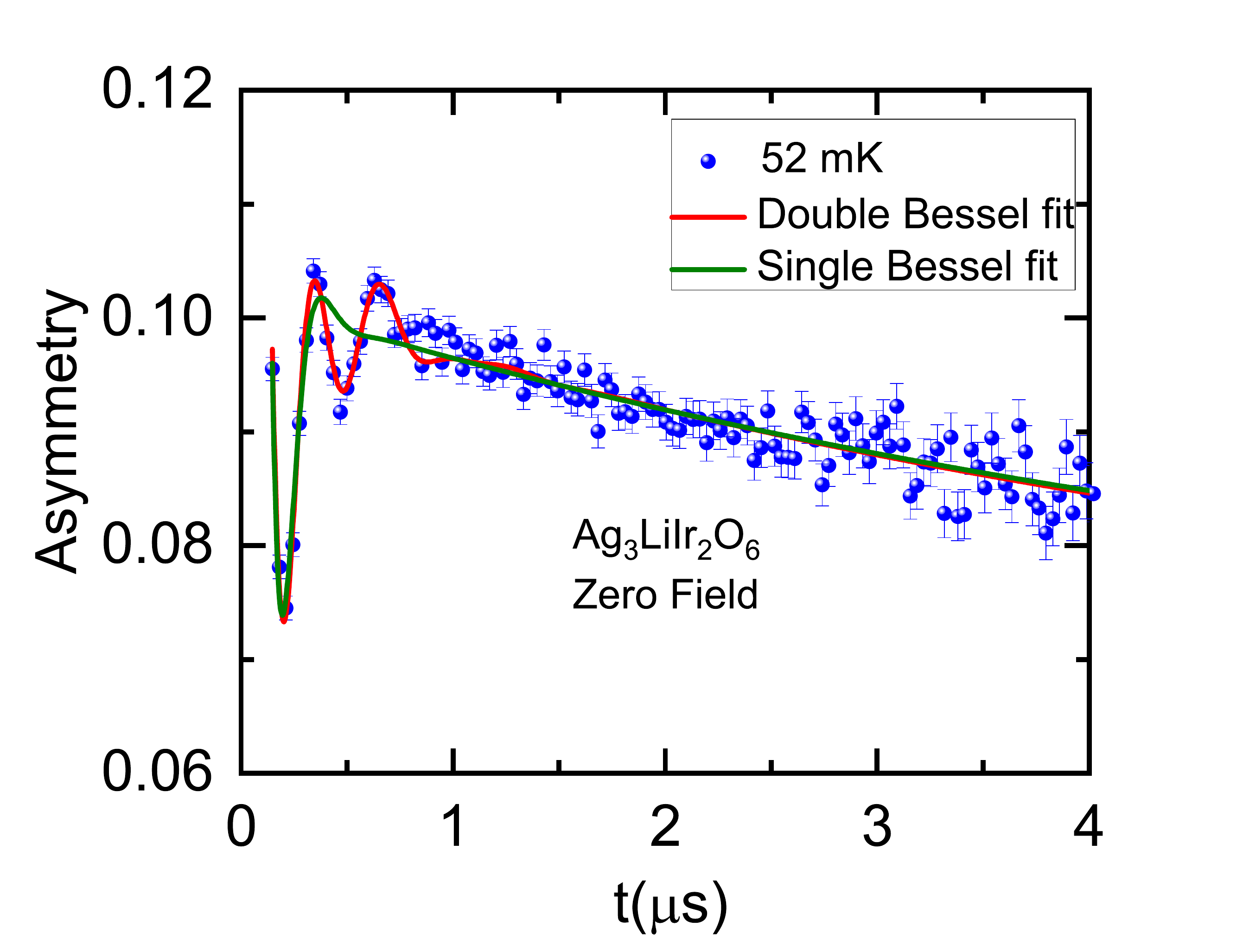}
	\caption{Variation of the muon asymmetry with time is shown at 52 mK together with a fit to both single and double Bessel functions.}
	\label{MuSR_BesselBoth}
\end{figure}

An introduction to the $\mu$SR technique can be found in Refs.~\onlinecite{Blundell_1999,Yaouanc2011}. As mentioned in the main text, clear oscillations were found in the zero field muon time spectrum below about 9 K.  Attempts to fit the data to one or two sinusoidal functions modulated by  exponentials did not lead to satisfactory fits. We then tried to fit the data using a Bessel function multiplied by an exponential (in addition, another exponential function was used to account for muons which have their magnetic moment parallel to the internal field).  Finally, we found that two Bessel functions were essential to obtain a satisfactory fit. A comparison of the fits to single and double Bessel functions is shown in Fig.~\ref{MuSR_BesselBoth}.

\newpage
\section{Heat Capacity}   
\begin{figure}[h!]
	\includegraphics[width=1.0\columnwidth]{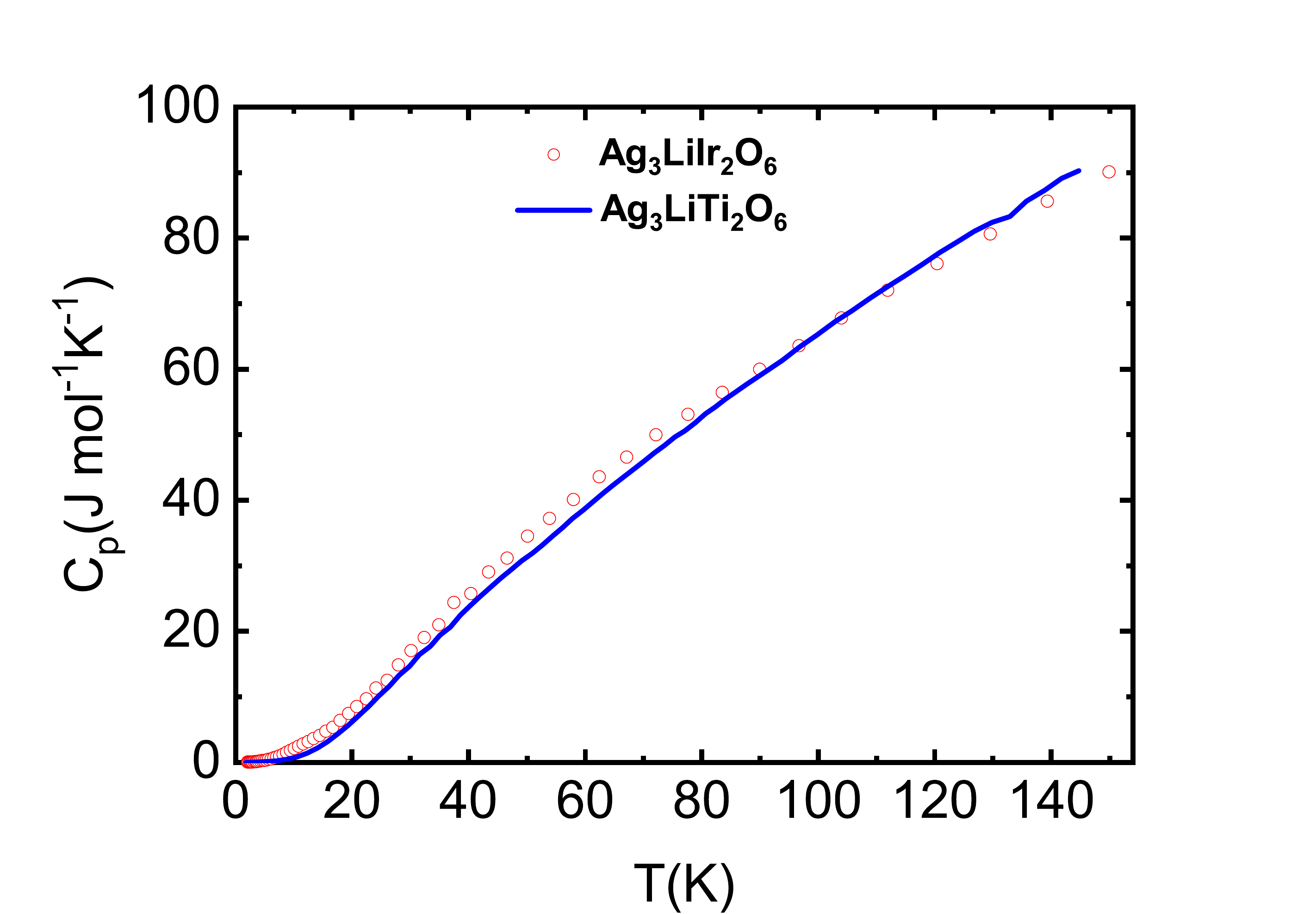}
	\caption{Heat capacity is plotted as a function of temperature in Ag$_3$LiIr$_2$O$_6$ (red open circles) and its nonmagnetic analog, Ag$_3$LiTi$_2$O$_6$ (blue solid line).}
	\label{heatcapacity}
\end{figure}

The heat capacity was measured for ALTO in the temperature range 2 K to 163 K.  The data for ALIO and ALTO are shown in Fig. \ref{heatcapacity}. As there is a molar mass difference and a unit cell volume difference between ALIO and ALTO, the lattice contribution  will be different for the two systems. A Bouvier scaling procedure \cite{Bouvier_1991} is often used to obtain a correction factor to scale the Debye temperature of the  nonmagnetic compound.  Correcting the temperature axis of the nonmagnetic compound by this factor then enables one to subtract the heat capacity of the nonmagnetic compound from the magnetic one.  In the present case, Bouvier scaling  gives the scaling factor $r=\theta_{Ir}/\theta_{Ti}=0.78(6)$ where $\theta_{Ir}$ and $\theta_{Ti}$ are the Debye temperatures of ALIO and ALTO, respectively. However, this did not lead to a complete cancellation of the heat capacity in the high-$T$ range (where the magnetic contribution is expected to be zero) when the data of the nonmagnetic compound ALTO (after temperature scaling) was subtracted from the magnetic compound ALIO. Finally, we used $r=\theta_{Ir}/\theta_{Ti}=0.88(7)$  to correct the temperature axis of ALTO in Fig.\ref{heatcapacity} which gives a better cancellation in the high temperature region. These data were subtracted from the ALIO data to obtain the magnetic heat capacity of  ALIO.

\newpage
\section{Magnetization}
\begin{figure}[h!]
	\includegraphics[width=1.0\columnwidth]{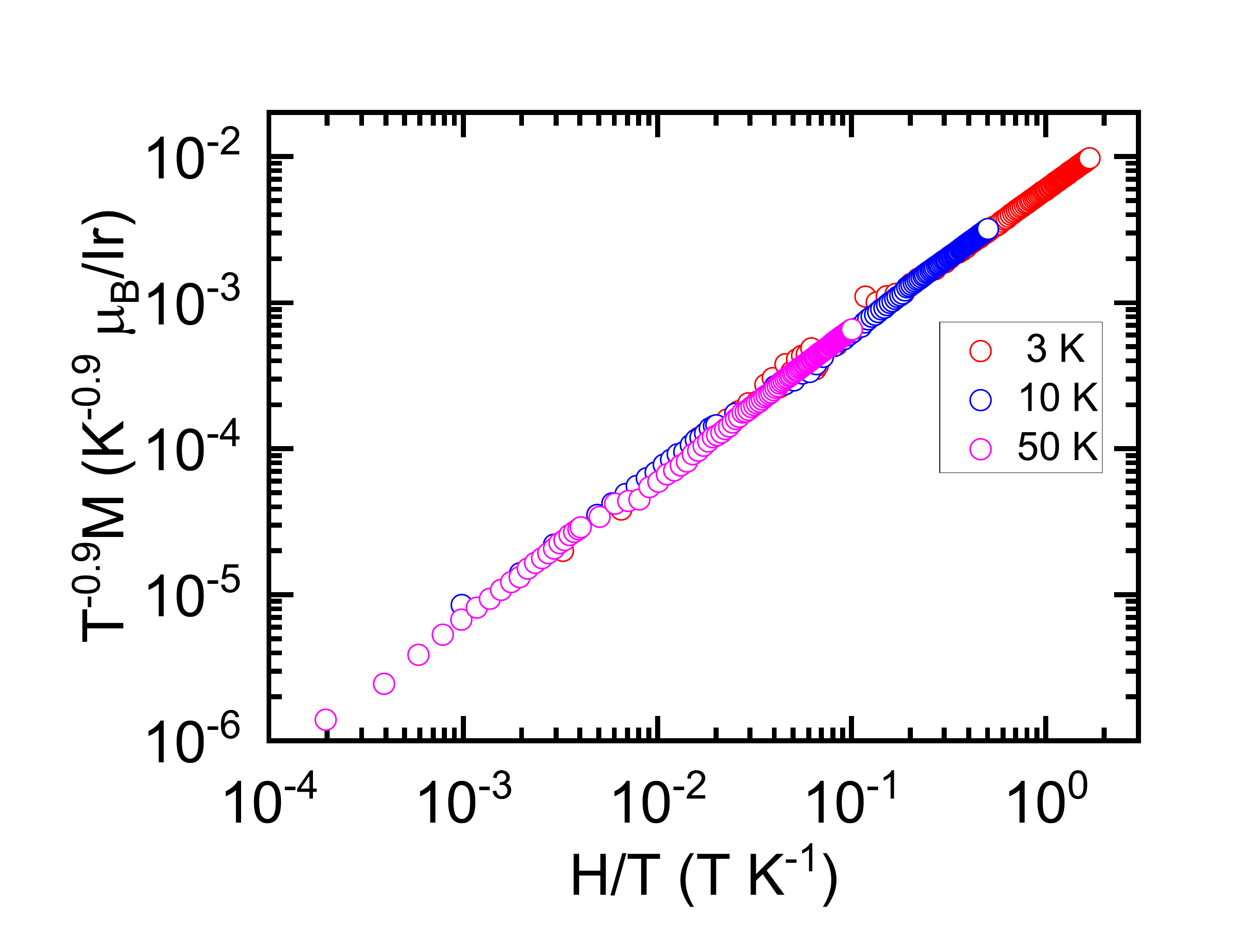}
	\caption{Data Collapse for T$^{\alpha-1}M$ as a function of $H/T$ for ALIO.}
	\label{MagnetisationScaling}
\end{figure}

We checked our $M(H, T)$ data for data collapse when $T^{\alpha -1 }M$ was plotted as a function of $H/T$.  As shown in Fig. \ref{MagnetisationScaling} our data exhibit a data collapse for $\alpha = 0.1$.  The corresponding exponent $\alpha$ was found to be 0.17 in Ref. \cite{Bahrami_T_2019}.

We also performed neutron diffraction on ALIO at 300 K and 3 K where no additional Bragg peaks were seen at low-$T$.  Given the large absorption cross section that Ir has for neutrons together with the incommensurate and two types of order as also persistent fluctuations inferred from our $\mu$SR data analysis, the absence of additional Bragg peaks is not surprising.

\section{Computational Details}
In order to achieve the convergence of energy eigenvalues, the kinetic energy cut off of the plane wave basis of Vienna Ab initio 
Simulation Package (VASP) code \cite{Kresse_A_1993,Kresse_E_1996} was chosen to be 600~eV. The Brillouin-Zone integrations are performed  
with 8$\times$8$\times$6 grid of kpoints. As mentioned in main text the GGA+U \cite{Anisimov_B_1991} calculations
are done in Dudarev scheme \cite{Dudarev_E_1998} with $U_{eff}= 1.5$~eV. The symmetry protected ionic relaxation of the crystal structure 
has been carried out within VASP calculation using the conjugate-gradient algorithm until the Hellman-Feynman forces on each atom were 
less than the tolerance value of 0.01~eV/\AA. In order to ascertain the accuracy of our VASP calculations we have
also performed the electronic structure calculation using Stuttgart TB-LMTO-47 code \cite{Andersen_E_1984} 
with K-mesh $8 \times 8 \times 6$, based on tight binding linearized muffin-tin orbital (TB-LMTO)
method in the atomic sphere approximation (ASA). The space filling in the ASA is obtained
by inserting appropriate empty spheres in the interstitial regions. The hopping parameters as well as onsite energies of the low-energy 
tight-binding model retaining only the Ir-t$_{2g}$ states in the basis are obtained from the muffin-tin orbital
(MTO) based N th order MTO (NMTO) downfolding method\cite{Andersen_M_2000}. 
\subsection{COHP calculation}
\begin{figure}[h!]
   \includegraphics[width=0.4\columnwidth]{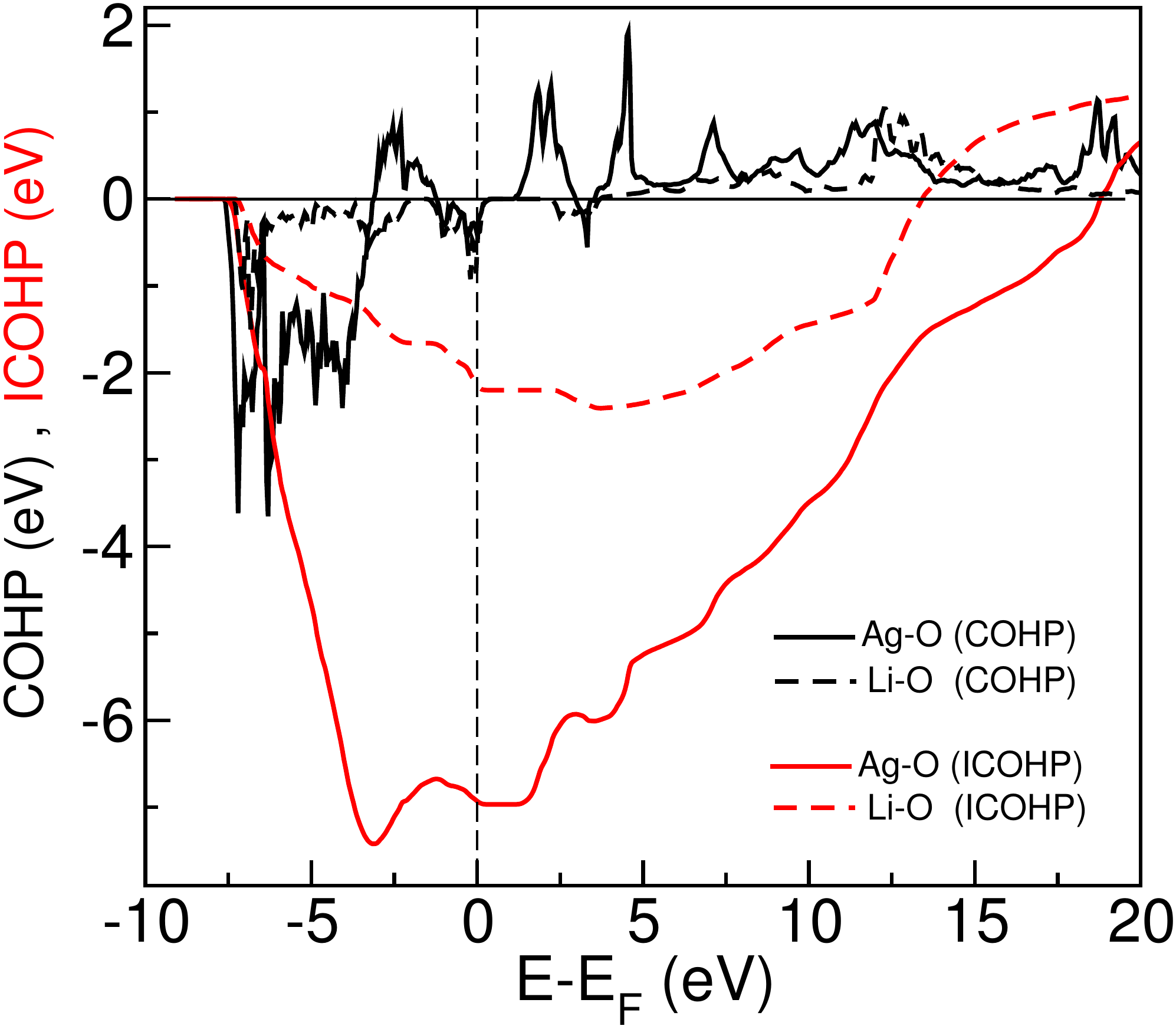}
\caption{COHP (black) and ICOHP (red) is plotted for bonds of Ag/Li atoms with O atoms for ALIO (solid line) and 
LIO (dotted line).  }
\label{cohpFig}
\end{figure}

The primary structural difference of ALIO with its parent is the enhanced interlayer 
distance due to the insertion of relatively large Ag atoms at the place of Li. In this regard the overlap between Ag-d and O-p
increases than Li-s O-p hybridization in $\alpha-$LIO. To get the quantitave estimation of hybridization we have plotted the 
an energy resolved visualization of the chemical bonding between the interlayer atoms with O which can be obtained from
the crystal orbital Hamiltonian population (COHP) plot as shown in Fig.\ref{cohpFig}. 
We have analyzed the chemical bonding by computing the crystal orbital Hamiltonian population (COHP) 
as implemented in the Stuttgart tight-binding linear muffin-tin orbital (TB-LMTO) code \cite{Andersen_E_1984}. The COHP
provides the information about the specific pairs of atoms  participating in the bonding, while the integrated COHP
(ICOHP) provides the strength of the interactions. The black and red solid (dotted) lines
 show the calculated COHP (ICOHP) of Ag-O and Li-O covalency for ALIO and $\alpha-$LIO respectively.
At E$_F$ the quantitave value of ICOHP of Ag-O is higher than that for Li-O bond in  $\alpha-$LIO which 
indicates that interlayer Ag-d is more hybridized with O that connects the superexchange path between nn Ir atoms within a-b 
plane. The electronic repulsion exerted on ligand O along the Ag-O linear bonds due to extended Ag-d orbitals 
increases the Ir-O-Ir bond angles which essentially enhances the Kitaev interaction between nn Ir atoms.
\subsection{Crystal Field and Hopping Integrals}
%%%%%%%%%%%%%%%%%%%%%%%%%%%%%%%%%%%%%%%%%%%%%%%%%%%%%%%%%
%\begin{figure}[h!]
%   \includegraphics[width=0.7\columnwidth]{diff_neighbor_si.eps}
%\caption{(a), (b) and (c) show $2^{nd}$, $3^{rd}$ and $4^{th}$ neighbor Ir-Ir exchange paths.}
%\label{neighborFig}
%\end{figure}
%%%%%%%%%%%%%%%%%%%%%%%%%%%%%%%%%%%%%%%%%%%%%%
In order to find the crystal-field splitting of Ir-d orbitals in the distorted octahedral environment and hopping interactions
between various Ir atoms, the NMTO downfolding method \cite{Andersen_M_2000} is employed. To calculate crystal-field splitting,
only Ir-t$_{2g}$ orbitals are retained in the basis and the rest are downfolded. Due to rotation and distortions in the IrO$_{6}$
octahedral network t$_{2g}$ levels are primarily contributed by d$_{xy}$, d$_{3z^2-1}$ and d$_{xz}$ orbitals of Ir$^{4+}$
ions. In order to extract the low energy Hamiltonian within local frame of reference pointed along the Ir-O bond lengths
 (as shown in Fig.1(b) of main text) we have rotated the basis (Euler angles: $\alpha= 178.55^0, \beta= -70.92^0, \gamma= -46.03^0$).
 The monoclinic distortion of the octahedra completely lifts the degeneracy of the t$_{2g}$ levels with gap of $\sim 0.27$ eV and
 $\sim 0.29$ meV w.r.t the lowest energy level. 

\subsection{Spin-polarized calculation}
\begin{figure}[h!]
 \includegraphics[width=0.8\columnwidth]{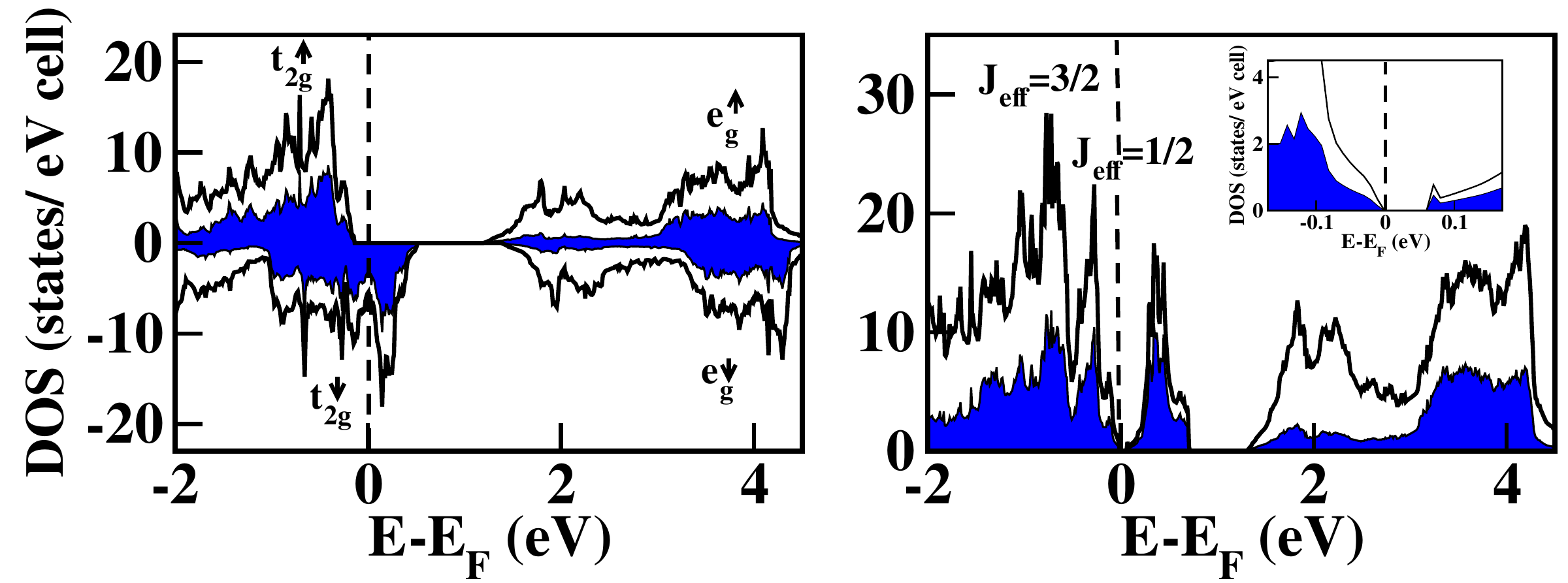}
\caption{Ferromagnetic total (black line) and partial Ir-d (blue shaded region) DOS plot within GGA+U (left panel)
	and GGA+SO+U (right panel) calculations. The zoomed in view of the GGA+SO+U DOS is shown in the inset of right panel to show the insulating nature.   }
\label{irfmFig}
\end{figure}

As a representative spin polarized calculation for ALIO we have considered here GGA+U calculation with ferromagnetic (FM)  arrangement of Ir spins. 
A plot of the spin-polarized density of states (DOS) for the FM configuration is shown in the left panel of Fig.\ref{irfmFig}. Calculations
reveal that in the majority spin channel the Ir-d t$_{2g}$ states are completely filled while as expected the
minority spin channel is partially occupied. The  magnetic moment per Ir site is calculated to be $0.63 \mu_B$, and the
rest of the moment are found to be at the ligand sites ($0.10\mu_B$ /O) due to hybridization. Next we have introduced SOC in  our
calculation (see Fig.\ref{irfmFig}). The spin and the orbital
moments in the GGA+U+SOC scheme at the Ir site is calculated to be $0.19 \mu_B$ and $0.41 \mu_B$ respectively. This large value of orbital moment suggests that  spin
orbit coupling induces the pseudo-spin J$_{eff}=1/2$ ground state (right panel of see Fig.\ref{irfmFig})
for this system. Further incorporation of U in presence of SOC splits the two fold degenerate J$_{eff}=1/2$ state creating a gap of value $\sim$66 meV, which leads to the insulating nature of the system.

\subsection{Details of calculation of the muon stopping site}
In order to locate the muon stopping site, we have implanted a muon within the crystallographic unit cell in the scope of our DFT 
calculation by inserting H$^+$ ion. In oxides muons are generally found to stop approximately $1$ \AA~away from an oxygen atom, similar to hydrogen in a hydroxyl bond
\cite{Holzschuh_M_1983,Denison_M_1984}.
Starting from a grid of interstitial positions near oxygen, the effect of muon embedding is inspected. Hence we consider a sphere around 
an oxygen of radius greater than $1$ \AA~(here chosen $1.2$ \AA~for our calculations) and random possible points on the surface of the sphere 
are chosen as the initial guess for the position of the muon. There are six equivalent oxygen positions in the formula unit that are 
generated from two inequivalent oxygen positions. Hence we have 
repeated the excercise once for oxygen of type-1 then for oxygen type-2. The muon implanted unit cell geometry and lattice parameters, for 
each set of starting guess, are allowed to optimize within VASP calculation using the conjugate-gradient algorithm until the Hellman-Feynman 
forces on each atom are less than the tolerance value of $0.01$~eV/\AA. The muon implanted near type-2 oxygen gives  lower energies than 
that near type-1 oxygens. The optimal location of the muon that gives rise to the lowest energy is $r_\mu = (0.0043,~0.0658,~0.1918)$, in 
fractional co-ordinate and at a distance $0.996$ \AA~from the nearest type-2 oxygen. The presence of the muon distorts the cell geometry. 
After obtaining the muon stopping site from DFT calculations, the dipole field can be calculated for a given spin order of the magnetic atoms with the formula \cite{BLUNDELL2009581,Williams_2016},
\begin{equation}
B^p(\textbf{r})= \frac{\mu_0}{4\pi} \sum_{i,q} \frac{\mu_{eff,i} m^{q}_{i} }{R^3_i} \Big( \frac{3R^p_iR^q_i}{R_i^2}-\delta^{pq}\Big),
\end{equation}
where $p,~q$ run over $x$, $y$. $z$ directions, \textbf{\textit{r}}$_i$ is the position of the $i^{th}$ magnetic ion, within the Lorentz 
sphere centered at muon site, $r_\mu$. Here \textbf{\textit{R}}$_i\equiv$ \textbf{\textit{r}}$_\mu$-\textbf{\textit{r}}$_i$ with 
$R_i=|\textbf{\textit{R}}_i|$. The effective magnetic moment of $i^{th}$ magnetic Ir ion is mentioned as $\mu_{eff,i}$ and $m^q_i$ is the 
direction cosine of that moment along direction $q$. Magnetic fields at the muon site are calculated with dipole sums 
performed in real space with a Lorentz sphere radius of $120$ \AA. The chosen radius produced results at a muon stopping site which 
converged within $0.2$ Oe compared to those using spheres of radii $150$ \AA~and $200$ \AA. The components of the dipolar field calculated with 
$\mu_{eff,i} =0.5$ $\mu_B$ /Ir (as obtained within the GGA+SOC+U calculations) are $B^x= -126.93$ Oe, $B^y= 9.90$ and $B^z= 55.05$. 
The magnitude of the obtained field was found to be  $|B| \sim 139$ Oe for the  N\'{e}el phase. 
%Mapping the experimentally observed observed average dipolar field ($220$ Oe) 
%and the DFT obtained muon stopping sites in the formula gives the value of effective moment $\sim 0.79~\mu_B$ per magnetic atom. 
The value of effective magnetic moment obtained from our theoretical calculations agrees well with that of experimental mapping except slight variation
may be due to presence of finite moments at neighboring ligand sites or choice of $U$ values for Ir atoms in our DFT calculations. 
\par From the DFT calculations, we find N\'{e}el and stripy phases are  close in energy (energy difference $\sim 7$ meV/f.u.), hence we have 
calculated the magnitude of the dipolar field for the stripy phase as well with $\mu_{eff,i} =0.5$ $\mu_B$ /Ir and obtained a field strength 
$266$ Oe which is close to the experimentally observed second component around $230$~Oe.

\bibliography{ref_sm}